\documentclass[journal,12pt,onecolumn,draftclsnofoot,]{IEEEtran}
\usepackage{cite}
\usepackage{amsmath,amssymb,amsfonts}
\usepackage{algorithmic}
\usepackage{graphicx,color}
\usepackage{textcomp}
\def\BibTeX{{\rm B\kern-.05em{\sc i\kern-.025em b}\kern-.08em
    T\kern-.1667em\lower.7ex\hbox{E}\kern-.125emX}}
\AtBeginDocument{\definecolor{ojcolor}{cmyk}{0.93,0.59,0.15,0.02}}
\usepackage{xcolor}
\usepackage{cite}
\usepackage{amsmath,amssymb,amsfonts}
\usepackage{caption}
\usepackage[stable]{footmisc}
\usepackage{mathrsfs}
\usepackage[linesnumbered,ruled,vlined]{algorithm2e}
\usepackage{braket}
\usepackage{graphicx}
\usepackage{textcomp}
\usepackage{xspace}
\usepackage{amsthm}
\usepackage[nolist]{acronym}
\usepackage{hyperref}
\theoremstyle{plain}
\newtheorem{thm}{Theorem}
\newtheorem{lem}[thm]{Lemma}

\usepackage{subfig}
\usepackage{tabularx}
\usepackage{booktabs} 
\colorlet{shadecolor}{yellow}

\begin{document}

\bstctlcite{}
\title{An Unsupervised Machine Learning to Optimize Hybrid Quantum Noise Clusters for Gaussian Quantum Channel}

\author{Mouli~Chakraborty,~\IEEEmembership{Student Member,~IEEE,}
      Anshu~Mukherjee,~\IEEEmembership{Member,~IEEE,}\\
      Ioannis~Krikidis,~\IEEEmembership{Fellow,~IEEE,}
      Avishek~Nag,~\IEEEmembership{Senior Member,~IEEE,}
      and~Subhash~Chandra,~\IEEEmembership{Member,~IEEE}

  \thanks{This publication has emanated from research supported by a grant from Science Foundation Ireland under Grant number 18/CRT/6222. For Open Access, the author has applied a CC BY public copyright license to any Author Accepted Manuscript version arising from this submission. This publication has emerged from research supported partly by a grant from IEEE Antennas and Propagation Society Graduate Fellowship Program - Quantum Technologies Initiative under Grant number `IEEE Funding – 1421.9040008''}
  \thanks{M. Chakraborty and S. Chandra are with the School of Natural Science, Trinity College Dublin, The University of Dublin, College Green, Dublin 02, Ireland (e-mail: chakrabm@tcd.ie) Ireland (e-mail: chakrabm@tcd.ie; SCHANDRA@tcd.ie).}
  \thanks{A. Mukherjee is with the School of Electrical and Electronic Engineering, University College Dublin, Belfield, Dublin 04, Ireland (e-mail: anshu.mukherjee@ieee.org).}%
  \thanks{I. Krikidis is with the IRIDA Research Centre for Communication Technologies, Department of Electrical and Computer Engineering, University of Cyprus, 1678 Nicosia, Cyprus (e-mail: krikidis@ucy.ac.cy).}
  \thanks{A. Nag is with the School of Computer Science, University College Dublin, Belfield, Dublin 04, Ireland (e-mail: avishek.nag@ucd.ie).}
  }


\maketitle

\begin{abstract}
This work focuses on optimizing the hybrid quantum noise model to improve the capacity of Gaussian quantum channels using \ac{ML} generated clusters. The work specifically leverages \ac{GMM} and the \ac{EM} algorithm to model the complex noise characteristics of quantum channels. Hybrid quantum noise, which includes both quantum shot noise and classical \ac{AWGN}, is modeled as an infinite mixture of Gaussian distributions weighted by Poissonian parameters. The study proposes a method to reduce the number of clusters within this noise model, simplifying visualization and improving the accuracy of channel capacity estimations without compromising essential noise characteristics. Key contributions include the reduction of Gaussian clusters while maintaining error tolerances and using the \ac{EM} algorithm to update quantum channel parameters, leading to more accurate channel capacity. The approach is validated through simulations, demonstrating that \ac{ML}-enhanced quantum noise clustering significantly improves the channel's performance in satellite-based quantum communication systems, specifically for \ac{QKD}. The work demonstrates  that \ac{GMM} and \ac{EM} algorithms provide a practical solution for modeling quantum noise in real-time applications, advancing the optimization of quantum communication networks.

\end{abstract}


\maketitle
\section{INTRODUCTION}
\IEEEPARstart{Q}{uantum} communication offers unmatched security and effectiveness in information transmission, addressing crucial data and information security issues. By transferring quantum information through quantum channels, utilizing quantum-mechanical elements such as photons, quantum communication enables revolutionary functions, including quantum cryptography \cite{bennett2014quantum}, quantum teleportation \cite{gisin2002quantum}, and the establishment of quantum networks \cite{kimble2008quantum}. Despite its potential, quantum communication systems face significant challenges, particularly quantum noise, which arises from the interaction of quantum systems with their environment or during measurement processes. This noise disrupts the accuracy and reliability of quantum information transmission, degrading essential features like coherence \cite{imre2005quantum} and entanglement \cite{wilde2013quantum} critical for quantum communication. 


Quantum noise comes in various forms, including quantum Poisson and classical noise, such as \ac{AWGN} \cite{pirandola2018photonicquantumsensing}, \cite{haykin2001communication}. Poisson noise is inherent to quantum systems due to the discrete nature of light and matter, while \ac{AWGN} originates from thermal fluctuations and electronic components in classical systems \cite{fox2006quantum}. Managing these noise sources effectively is essential for maintaining the integrity of quantum communication channels, and hybrid noise models, combining both quantum and classical noise, are crucial for accurately capturing the full range of disturbances that affect such systems \cite{mouli2024,chakraborty2024hybridquantumnoiseapproximation}. These models enable more realistic assessments of \ac{QKD}, error rates, information capacity, and security, allowing for optimizing quantum technologies \cite{Mouli2024Asymp_QKD_SatComm,Mouli2024Finite_size_QKD_QComm},\cite{Gyongyosi2018}. Optical cluster states \cite{Nielsen_Optical_cluster_states} are critical in achieving quantum computational universality in \ac{LOQC} \cite{kok2007LOQC}. Since establishing direct entangling operations between photons is challenging due to the need for nonlinear effects, the probabilistic generation of entangled resource states offers a promising alternative approach. In \ac{LOQC}, cluster states can be created using different methods, depending on how quantum information is encoded on a silicon photonic chip. Standard encoding options include spatial modes, where quantum information is encoded in photons' paths, and polarization, where the quantum state is defined by the orientation of the photons’ electric fields \cite{langford2007encoding}. The specific method of generating cluster states depends on the chosen encoding, which is vital for effectively operating quantum computing tasks. These states are essential building blocks for performing quantum gates \cite{Nielsen_Optical_cluster_states} and operations in \ac{LOQC} systems, driving advancements in photonic quantum computing.

\ac{ML} methods could have a crucial impact on improving the estimation and optimization of quantum channel parameters \cite{mafu2024MLQCOM}, error correction \cite{liao2023ML_QErrorCorrection}, resource allocation \cite{duong2022ML_Q_ResourceAllocations}, and quantum information transmission. Progress has been notable in quantum communication technology for free-space satellites \cite{vallone2015experimental}. However, challenges related to quantum repeaters for long-distance quantum communications still need to be addressed to establish a global quantum network \cite{wallnofer2020ML_Q_repeaters},\cite{simon2017global_quantum_network}. Free-space quantum communication channels have encountered difficulties such as loss in the quantum channel, security issues, and limited data rates \cite{liao2018satellite_quantum_network}. Conventional \ac{ML} techniques offer a valuable approach to characterizing the features of free-space quantum channels \cite{djordjevic2020global_Q_networks}. In a recent study \cite{ismail2019freespace}, supervised \ac{ML} was employed to forecast the atmospheric strength of a free-space quantum channel. \ac{ML} techniques play a crucial role in addressing quantum noise in communication systems. Clustering methods like \ac{GMM}\cite{chatzis2012_GMM}, which represent noise as a mixture of Gaussian components, offer flexible and accurate noise modeling. The \ac{EM} algorithm \cite{mclachlan2007EMalgo} calculating the \ac{MAP}
\cite{murphy2012machine}, used with \ac{GMM}, refines the model parameters iteratively to optimize noise clustering, thereby improving quantum channel capacity and noise management. This approach helps manage the probabilistic nature of quantum noise, enhances error correction, and boosts the precision of quantum communication protocols \cite{djordjevic2012quantum}. The \ac{EM} algorithm  is particularly useful for modeling hybrid quantum noise in practical channels, including satellite-based \ac{QKD} systems \cite{Mouli2024Asymp_QKD_SatComm, Mouli2024Finite_size_QKD_QComm}. It efficiently processes high-dimensional quantum data and optimizes parameters like means, variances, and weights in noise clusters, balancing model complexity with computational feasibility. This makes \ac{EM} clustering essential for real-time quantum noise modeling and enhances the performance and reliability of quantum communication systems.

 {The proposed quantum channel optimization is highly effective for free-space optical satellite quantum communication, particularly in quantum key distribution (\ac{QKD}). Prior work \cite{Mouli2024Asymp_QKD_SatComm, Mouli2024Finite_size_QKD_QComm} has demonstrated that optimizing quantum noise clustering techniques and leveraging \ac{ML}-optimized parameters significantly enhance quantum channel capacity and \ac{QKD} rates, highlighting the practical relevance of this approach. An initial comparison with the K-Means algorithm was conducted, but the study primarily focused on the \ac{GMM}-based \ac{EM} algorithm for visualizing clustered datasets. Detailed analyses of clustering techniques, including the Elbow Method, Silhouette Analysis, and Gap Statistic, along with probabilistic approaches such as \ac{BIC} and \ac{AIC}, are discussed, emphasizing the evaluation of clustering quality and optimal cluster determination. The \ac{GMM} framework, supported by the \ac{EM} algorithm, was selected for its ability to iteratively optimize likelihood, handle missing data, and model overlapping clusters through soft probabilistic assignments. This makes it ideal for analyzing soft clustering in the hybrid quantum noise framework. Future work will explore \ac{DL}-based clustering methods and hard clustering approaches to extend this model's applicability. The current study assumes a bi-variate hybrid quantum noise dataset, aligning with the system's specifications.}

Our work proposes a \ac{ML} approach for visualization of hybrid quantum noise in higher dimensions. Our previously published work developed the hybrid quantum noise model and visualized it in lower dimensions. This research works to validate that model and visualise the hybrid quantum noise in $2$-dimension. Our previous works \cite{mouli2024,chakraborty2024hybridquantumnoiseapproximation} expressed the hybrid quantum noise model as \ac{GMM}, an infinite mixture of Gaussians with weightage from Poissonian distribution. It has been investigated as an intermediate result that the good approximation of the p.d.f. of the quantum noise depends on the interrelation of the Poisson parameter and the number of Gaussian components \cite{chakraborty2024hybridquantumnoiseapproximation}. In essence, the critical contributions of this study can be summarized in three main aspects: 
\begin{itemize}
     \item We introduced a method that reduces the number of Gaussian clusters and simplifies the visualization of hybrid quantum noise in higher dimensions. This method achieves up to $15\%$  error tolerance in Poisson weightage, ensuring accuracy  $85\%$  and making the visualization more feasible.
     \item The research demonstrates that manual input for initial channel parameters fails to identify the correct clusters, as seen in Fig. ~\ref{fig ground truth and initial guess}. In contrast, \ac{ML} -optimized parameters reveal the actual hybrid quantum noise clusters, as illustrated in Fig. ~\ref{fig Final it20 and log_likelihood_2_5}, helping uncover the hidden clusters within the noise.
     \item The reduction of clusters, along with applying the \ac{EM} algorithm, improves quantum channel capacity  $11.8\%$ without compromising essential hybrid noise components, leading to a more accurate capacity estimation, especially at varying SNR levels.
 \end{itemize} 

Our study focuses on integrating \ac{ML} techniques to model and optimize hybrid quantum noise in quantum communication systems. We reduce the noise model's complexity by utilizing \ac{GMM} and \ac{EM} algorithms, transforming an infinite mixture of Gaussians into a manageable finite model. This simplification allows for better visualization of quantum noise in higher dimensions and enhances the accuracy of quantum channel capacity estimation. Furthermore, this approach can be applied to real-world quantum networks, such as satellite-based \ac{QKD} systems, to improve secure communication.
Please note that a comprehensive list of the notations used throughout this paper is provided below. 

\textit{Notation:} We use $tr$ denote the trace of a matrix, and $\mathbf{T(\cdot)}$ represent a trace-preserving map $\mathbf{T}$. The adjoint of a matrix $\mathbf{A}$ is denoted by $\mathbf{A}^\dag$, while its transpose is represented as $\mathbf{A}^t$. The complex conjugate of a vector $\nu$ is written as $\nu^*$. The symbol $\otimes$ is used to denote the tensor product and $\mathbb{C}$ represents the set of complex numbers. The parameter $\lambda$ indicates the Poisson parameter and $\mathcal{N}$ is used for Gaussian density. Quantum states are represented using Dirac notation, where $\ket{\cdot}$ denotes a ket and $\bra{\cdot}$ denotes a bra. The notation $|\cdot|$ represents the norm.

\section{System Model for the Quantum Communication Channel}\label{System Model}


\subsection{Gaussian Quantum Channel Model} 
\label{Gaussian Quantum Channel equation}

The \ac{CV} Gaussian quantum channel model used here is adapted from our prior work \cite{chakraborty2024hybridquantumnoiseapproximation, mouli2024, Mouli2024Asymp_QKD_SatComm}. For detailed insights, readers are referred to those studies. A qubit is a superposition of basis states $\ket{\boldsymbol{0}}$ and $\ket{\boldsymbol{1}}$, expressed as $\ket{\boldsymbol{\psi}} = \alpha\ket{\boldsymbol{0}} + \beta\ket{\boldsymbol{1}}$, with $|\alpha|^2 + |\beta|^2 = 1$. Density matrices describe pure states 
	$\boldsymbol{\rho}=\ket{\boldsymbol{\psi}}\bra{\boldsymbol{\psi}}= \begin{bmatrix}
		|\alpha|^2 & \alpha^*\beta\\
		\alpha\beta^* &  |\beta|^2
	\end{bmatrix}$, 
where $\alpha^*$ and  $\beta^*$ are the complex conjugates of $\alpha$ and $\beta$, respectively., while mixed states are probabilistic combinations of pure states $\boldsymbol{\rho} = \sum_i p_i \ket{\boldsymbol{\psi_i}}\bra{\boldsymbol{\psi_i}}$, where $\sum_i p_i = 1$.


A quantum channel is a \ac{CPTP} map transforming density operators \(\boldsymbol{\rho}\) on a Hilbert space \(\mathcal{H}\) into other states \cite{mouli2024,chakraborty2024hybridquantumnoiseapproximation, cerf2007quantum, demoen1977completely, holevo1999capacity}, represented as \(\boldsymbol{\rho} \mapsto \mathbf{T}(\boldsymbol{\rho})\). Assuming identical input and output spaces, a channel \(\mathbf{T}\) can be expressed as \(\mathbf{T}(\boldsymbol{\rho}) = \text{tr}_E[\mathbf{U}(\boldsymbol{\rho} \otimes \boldsymbol{\rho}_E)\mathbf{U}^\dagger]\), where \(\mathbf{U}\) is a Gaussian unitary determined by a quadratic Bosonic Hamiltonian, and \(\boldsymbol{\rho}_E\) is a Gaussian state \cite{eisert2005gaussian, cerf2007quantum, mouli2024}. Gaussian channels include lossless unitary evolution governed by \(\mathbf{U} = e^{i/2}\sum_{k, l} \mathbf{H}_{kl} \mathcal{R}_k \mathcal{R}_l\), where \(\mathbf{H}_{kl}\) is a real symmetric matrix, \(\mathcal{R}\) are canonical coordinates, and \(\mathbf{U}\) represents the symplectic group \(Sp(2n, \mathbb{R})\) \cite{eisert2005gaussian, cerf2007quantum}. Channels are described in the Schrödinger picture as \(\boldsymbol{\rho} \mapsto \mathbf{T}_{\mathbf{A, Z}}(\boldsymbol{\rho})\) or via state transformations \(\boldsymbol{\rho} \mapsto {\mathbf{A}^T}\boldsymbol{\rho}\mathbf{A} + \mathbf{Z}\) \cite{eisert2005gaussian, cerf2007quantum, mouli2024, chakraborty2024hybridquantumnoiseapproximation, Mouli2024Asymp_QKD_SatComm}.

\begin{figure*}
	\centerline{\includegraphics[width = 1\textwidth]{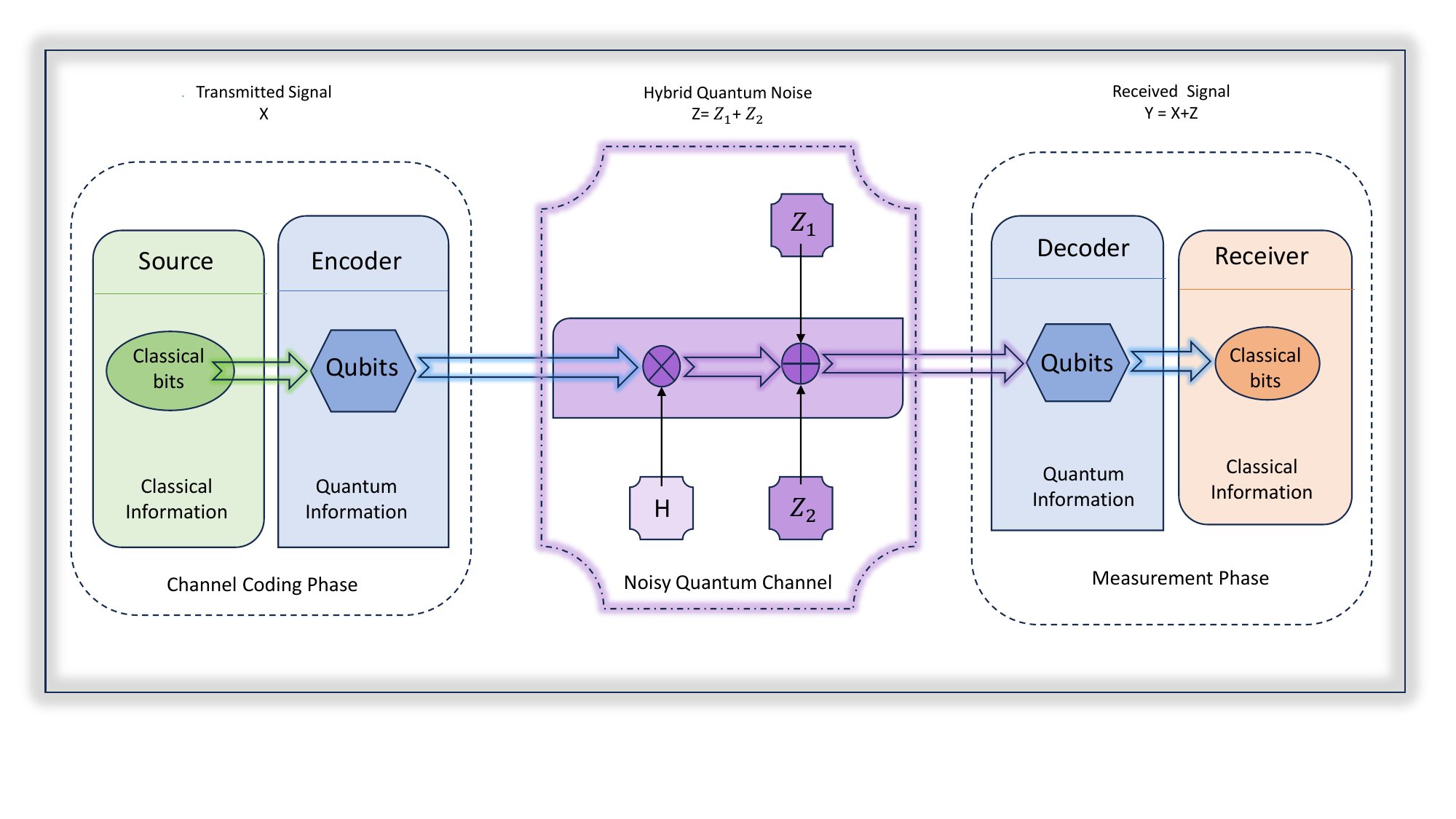}}
	\caption{A noisy Quantum Channel: A schematic diagram showing a single quantum link between transmitter and receiver. $H$ is the channel matrix, $Z$ is the hybrid quantum noise as a convolution of and quantum Poissonian noise $Z_1$ and additive white Gaussian noise $Z_2$. The complete received signal can be modeled using the channel equation $Y=X+Z$ , where $X$ is the transmitted signal. }
	\label{quantum noisy channel diag}
\end{figure*}

A pure-state qubit can be visualized as a point \((\theta, \phi)\) on the Bloch sphere's surface \cite{imreGyongyosi2012advancedQ_BOOK}, or equivalently described by a bivariate function \(\zeta(\theta, \phi)\). Under restricted noise power, the qubit remains within a strip on the Bloch sphere, centered on a circular path traced by varying \(\theta\) while keeping \(\phi\) constant. The strip's width is \(2\delta\), where \(\delta\) is the distance from the circle to the strip's edges, ensuring \((\theta, \phi) \mapsto (\tilde{\theta}, \phi \pm \delta)\) with \(\delta\) small and \(\phi\) nearly constant \cite{mouli2024,chakraborty2024hybridquantumnoiseapproximation}. For mixed-state qubits, represented as \((\theta, \phi, r)\) inside the Bloch sphere (\(r\) being the distance from the center), they can be modeled using a multivariate function \(\tilde{\zeta}(\theta, \phi, r)\). With noise, the qubit's position resides within a torus-like structure with a square cross-section, leading to \((\theta, \phi, r) \mapsto (\tilde{\theta}, \phi \pm \delta, r \pm \delta)\), where \(\delta\) is small, keeping \(\phi\) and \(r\) nearly constant relative to \(\theta\). This simplifies to \((\theta, \phi, r) \mapsto (\tilde{\theta}, \phi, r)\), approximating qubit behavior by varying one parameter while holding others constant. In both cases, a qubit \(\boldsymbol{\rho}\) can ultimately be described by a scalar \(\rho\) \cite{mouli2024,chakraborty2024hybridquantumnoiseapproximation,Mouli2024Asymp_QKD_SatComm}.


Let's return to the expression of a Gaussian channel, 
\(
\boldsymbol{\rho} \mapsto \mathbf{T}_{\mathbf{A, Z}}(\boldsymbol{\rho})
\),  where \(\mathbf{A}\) functions to modulate the signal through amplification, attenuation, and rotational adjustments in phase space. Meanwhile, \(\mathbf{Z}\) adds a noise component, which includes both quantum noise (necessary to render the transformation physically meaningful) and classical noise. Intriguingly, \(\mathbf{A}\) can be any real matrix, allowing any transformation \(\rho \mapsto \mathbf{A}^T \rho \mathbf{A}\) to be approximated, provided that adequate noise is incorporated. As detailed, a simple quantum channel can be modeled as \(\rho \mapsto \rho + \mathbf{Z}\), where \(\mathbf{Z}\) includes both quantum and classical components of noise \cite{mouli2024,chakraborty2024hybridquantumnoiseapproximation}. Consider, \(\mathbf{A = I}_{2n}\) where $\mathbf{I}_{2n}$ is the identity matrix of order $2n \times 2n$ and $n$ is the canonical degrees of freedom of the quantum system. \(\mathbf{A}\) can be any real matrix; the transformation is valid as a quantum channel if enough noise is introduced \cite{cerf2007quantum}. In such a scenario, \(\mathbf{A}^T \mathbf{\rho A}\) becomes scalar. The noise \(\mathbf{Z}\) added to the system is also treated as scalar from the convolution of \ac{AWGN} and quantum Poisson noise with both components of the noises in scalar representation \cite{mouli2024,chakraborty2024hybridquantumnoiseapproximation}. Hence, this setup can be equated to the basic formula in communication theory where \(Y = X + Z\), which represents the transmitted signal \(X\), the received signal \(Y\), and the hybrid quantum noise \(Z\) impacting the channel from various sources. In a practical quantum communication network handling classical information, it is presumed that the links are influenced by both quantum and classical noise. This noise is assumed to be additive, where \(Z = Z_1 + Z_2\), with \(Z_1\) being Poisson-distributed quantum shot noise and \(Z_2\) representing Gaussian-distributed white classical noise \cite{mouli2024,chakraborty2024hybridquantumnoiseapproximation}. This concept is further illustrated in the referenced  Fig.~\ref{quantum noisy channel diag}, which provides a schematic overview of the theory.

\subsection{Hybrid Noise model}
For quantum communication, it is essential to model the realistic quantum noise of a quantum channel. Previous studies show that quantum noise is widely represented as Poissonian noise \cite{Paul1982photon,renker2009advances, eleftheriadou2013quantum}. However, a practical quantum link also suffers from noise sources that contribute non-quantum noises, such as classical noise sources \cite{Johnson1928, Leeson1966, Voss1979}. Classical noise is the integrated part of the quantum channel and must be addressed to model the realistic quantum channel. 
In the semi-classical theory of photodetection, the shot noise level corresponding to Poissonian photoelectron statistics is the fundamental detection limit. Hence, the shot noise power level is often called the quantum limit or the standard quantum limit of detection. Similarly, shot noise is often called quantum noise \cite{cerf2007quantum, fox2006quantum}. Considering the qubit noise as Poissonian and the classical noise as additive white Gaussian, one can expect a hybrid quantum noise represented by the following \ac{p.d.f.} \cite{mouli2024,chakraborty2024hybridquantumnoiseapproximation},

\begin{equation}
	f_{Z}(z) 
	= \sum_{i=0}^{\infty}
	\frac{e^{-\lambda}\lambda^i}{i!}\frac{1}{\sigma_{Z_2}\sqrt{2\pi}}e^{-\frac{1}{2}\Big(\frac{z-i-\mu_{Z_2}}{\sigma_{Z_2}}\Big)^2}.\\
	\label{eq The pdf of hybrid quantum noise} 
\end{equation} 

\subsection{Quantum Transmitted Signal Model}

For the transmitted signal model, let us consider a Gaussian distributed signal for the \ac{CV} system. Mathematically, a state is Gaussian if its distribution function in phase or its density operator in the Fock space is Gaussian \cite{mouli2024}. Examples of Gaussian functions are well-known \ac{p.d.f.} of Normal distribution, Wigner function, etc. Important quantum information processing experiments are done with quantum light, described as Gaussian distributed \cite{cerf2007quantum, fox2006quantum}. In our model, the input quantum signal $X$ in terms of qubits is in a Gaussian state, and it can be characterized by the scalar random variable $X$ having a Gaussian distribution with parameters $\mu_{X}$ and $\sigma_{X}$, 
\(X~ \sim \mathcal{N} (x;\mu_{X},\,\sigma_{X}^{2}) \)
the \ac{p.d.f.} $f_{X}$ of $X$ can be expressed as  \cite{mouli2024,chakraborty2024hybridquantumnoiseapproximation},
 $f_{X}(x)=\frac{1}{\sigma_{X}\sqrt{2\pi}}e^{-\frac{1}{2}\big(\frac{x-\mu_{X}}{\sigma_{X}}\big)^2 }$,
where $\mu_{X}$ and $\sigma_{X}$ are the distribution's mean and standard deviation (s.d.).

\section{Methodology of Quantum Noise Cluster Classification for \\ Quantum Communication }


\begin{figure*}
\centerline{\includegraphics[width = 1\textwidth]{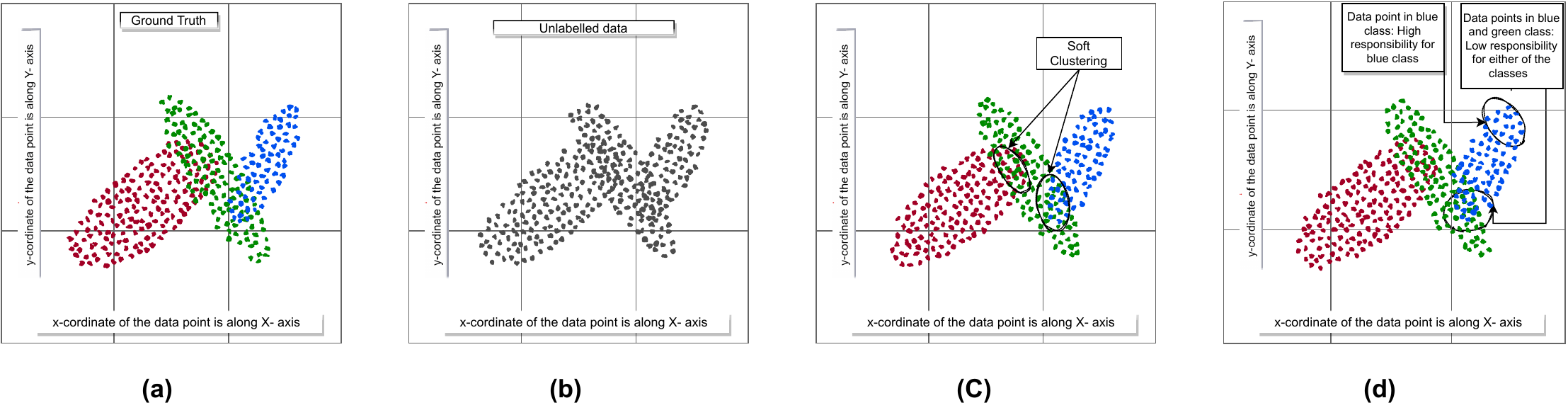}}
\caption{A generalized categorization of hybrid Quantum Noise data in two dimension space: (a) Original data (termed as Ground Truth) with specific cluster; (b) Unlabeled data for finding the original clusters; (c) \ac{GMM} used for soft-clustering; (d) Responsibility of a cluster. Note that the x-coordinates and y-coordinates of the bi-variate data points are considered along the graphs' X-axis and Y-axis. }
\label{Gaussian Mixture Model used for soft-clustering with Responsibility of a cluster}
\end{figure*}

\subsection{Clustering with Gaussian Mixture Model} \label{IIIA}

Let us start with the hybrid quantum noise model in \eqref{eq The pdf of hybrid quantum noise},
\begin{equation}
\begin{split}
	f_{Z}(z) & 
	= \sum_{i=0}^{\infty} w_{i}.\mathcal{N} (z;\mu_{i}^{(z)},\,{\sigma_{i}^{(z)}}^{2}),
	\label{eq The pdf of joint quantum noise} 
\end{split}
\end{equation} 
where $w_{i}=\frac{e^{-\lambda}\lambda^i}{i!}$ represent the weightage of the mixtures of Gaussians, $\sum_{i=0}^{\infty} w_{i}=1$, $w_{i}\geq 0 \quad  \forall i$ and $\mathcal{N} (z;\mu_{i}^{(z)},\,{\sigma_{i}^{(z)}}^{2}) $ is the Gaussian density in random variable $Z$, mean $\mu_{i}^{(z)}=\mu_{Z_2}+i$ and ${\sigma_{i}^{(z)}}^{2}={\sigma_{Z_2}}^2$ with standard deviation $\sigma_{i}^{(z)}$ and variance ${\sigma_{i}^{(z)}}^{2}$. These lead to visualizing the noise model data in a scalar space or lower dimensions. In the meantime, this infinite mixture can be approximated by a finite number of components as 
\begin{equation}
\begin{split}
 f_{Z}(z) 
= \sum_{i=0}^{R} w_{i}.\mathcal{N}\Big(z;\mu_{i}^{(z)},\,{\sigma_{i}^{(z)}}^{2}\Big).
\label{eq approx pdf of joint quantum noise in Gaussian form} 
\end{split}    
\end{equation}
This is the Gaussian mixture in the scalar variable $Z$, $w_{i}\geq 0 \quad  \forall i$ and $\sum_{i=0}^{R} w_{i} \approx 1$ for large $R$ it depends on Poisson parameter $\lambda$. However, a higher-dimensional approach is needed to visualize the original hybrid quantum noise data in the corresponding vector space as a scatter plot. This leads to the noise\ac{p.d.f.}in a higher dimension given by the corresponding Gaussian mixture in a random vector $\mathbf{z} \in \mathbb{R}^D$ as

\begin{equation}
f_{\boldsymbol{Z}}(\mathbf{z}) = \sum_{i=0}^{R} w_{i}.\mathcal{N} \Big(\mathbf{z};\boldsymbol{\mu}_{i}^{(z)},\,{\boldsymbol{\Sigma}_{i}^{(z)}}\Big),
\label{eq approx pdf of joint quantum noise in Gaussian form for vector}   
\end{equation}
where $w_{i}=\frac{e^{-\lambda}\lambda^i}{i!}$ represents the weightage of the Gaussian mixtures, $\mathbf{z}$ is the random vector in $\mathbb{R}^D$, $D$ denotes the dimension of the vector $\mathbf{z}$, $\boldsymbol{\mu}_{i}^{(z)}$ is the mean vector, $\boldsymbol{{\Sigma}}_{i}^{(z)}$ is the covariance matrix of the corresponding Gaussian density $\mathcal{N}$, and the multivariate Gaussian is given by 
 $ \ensuremath{\mathcal{N}(\mathbf{z}_{n};\boldsymbol{\mu}_{k}^{(z)},\boldsymbol{\Sigma}_{k}^{(z)})}
\ensuremath{=\frac{1}{(2\pi)^{D/2}|\boldsymbol{\Sigma}_{k}^{(z)}|^{1/2}}e^{-\frac{1}{2}(\mathbf{z}_{n}-\boldsymbol{\mu}_{k}^{(z)})^{T}\{\boldsymbol{\Sigma}_{k}^{(z)}\}^{-1}(\mathbf{z}_{n}-\boldsymbol{\mu}_{k}^{(z)})}.}$  

Consider the hybrid quantum noise data set $\boldsymbol{Z}=\{\mathbf{z}_1,\mathbf{z}_2,...,\mathbf{z}_n\}$, $\mathbf{z}_i \in \mathbb{R}^D$, where $D$ denotes the data points dimension , respectively. The primary purpose of this work is to understand the multidimensional hybrid quantum noise data set by partitioning it into $K$ the number of clusters, maximizing the likelihood of the probabilistic model. The hybrid quantum noise model in \eqref{eq approx pdf of joint quantum noise in Gaussian form for vector} can be expressed as 
$f_{\boldsymbol{Z}}(\mathbf{z})=\sum_{t}^{}  f_{\boldsymbol{Z}}(\mathbf{z},t)= \sum_{t}{}f_{\boldsymbol{Z}}(\mathbf{z}|t) f_{\boldsymbol{Z}}(t)$,
where $t$ is the latent variable, approximating the distribution with a Gaussian mixture. $ f_{\boldsymbol{Z}}(\mathbf{z}|t)$ are Gaussian \ac{p.d.f.}, and $ f_{\boldsymbol{Z}}(t)$ are mixing coefficients, these coefficients giving the weightage to the distribution $ f_{\boldsymbol{Z}}(\mathbf{z}|t)$ with a particular probability of $ f_{\boldsymbol{Z}}(t)$. Plugging over the $ f_{\boldsymbol{Z}}(\mathbf{z}|t)$ with weightage $ f_{\boldsymbol{Z}}(t)$, the resultant sum $\sum_{t}{} f_{\boldsymbol{Z}}(\mathbf{z}|t) f_{\boldsymbol{Z}}(t)$ represents the \ac{GMM}, where $f_{\boldsymbol{Z}}(\mathbf{z}|t)$ is the posterior probabilities of the latent variable $t$ given an observable $\mathbf{z}$. From \eqref{eq approx pdf of joint quantum noise in Gaussian form for vector}, it guarantees the Gaussian distribution for each latent variable $t$'s conditional distribution $ f_{\boldsymbol{Z}}(\mathbf{z}|t)$.

To use the \ac{EM} algorithm for clustering components in the hybrid quantum noise dataset, a discrete random variable is introduced to assign data points to clusters (\( t \), the \( t \)-th component of the mixture), ensuring that each cluster satisfies a Gaussian distribution with its parametric values. Ground truth in \ac{ML} serves as the benchmark for training and evaluating models, ensuring accurate cluster assignments. In \ac{GMM}-based clustering, the probability of a quantum data point \( \mathbf{z} \) belonging to a specific cluster (e.g., red, green, or blue) is calculated. For classified data, points are color-coded by cluster, as shown in Fig.~\ref{Gaussian Mixture Model used for soft-clustering with Responsibility of a cluster}(a), while unclassified data is shown in Fig.~\ref{Gaussian Mixture Model used for soft-clustering with Responsibility of a cluster}(b). When cluster assignment is uncertain, overlapping probabilities are represented by black circles (soft clustering) in Fig.~\ref{Gaussian Mixture Model used for soft-clustering with Responsibility of a cluster}(c). In this probabilistic framework, each data point is assigned a responsibility, the probability of belonging to a cluster, as depicted in Fig.~\ref{Gaussian Mixture Model used for soft-clustering with Responsibility of a cluster}(d). \ac{GMM} excels in quantum-specific problems due to its soft probabilistic assignments (\([0,1]\)) compared to the hard classifications (\(0\) or \(1\)) of K-Means \cite{patel2020clustering}. The \ac{EM} algorithm is optimal for \ac{GMM}, as it is tailored for probabilistic models and handles Gaussian-distributed data efficiently. This makes \ac{GMM} and the \ac{EM} algorithm ideal for quantum data classification, where soft clustering is preferred.

\setlength{\parskip}{1pt}

 The \ac{p.d.f.} of the \ac{GMM}, which models the distribution of data points in a clustering scenario, is given by \eqref{eq approx pdf of joint quantum noise in Gaussian form for vector}. The discrete latent variable $t_k \in \{0,1\}$ for the $K$ clusters is represented by the prior \(f_{\boldsymbol{Z}}(t_k =1)=\frac{e^{-\lambda}\lambda^k}{k!},\) $k\in [0,K]$, $\sum_{k=1}^{K}\frac{e^{-\lambda}\lambda^i}{i!} \approx 1$ for a sufficient choice of $K$. The clusters of the underlying components are Gaussians, with different parameters 
   $f_{\boldsymbol{Z}}(\mathbf{z}|t_k=1)=\mathcal{N} \Big(\mathbf{z};\boldsymbol{\mu}_{k}^{(z)},\,{\boldsymbol{\Sigma}_{k}^{(z)}}\Big)$.
The quantity \(t_k \in \{0,1\}\) indicates whether the data point \(\mathbf{z}\) belongs to the \(k\)-th class or not. Specifically, \(t_k=0\) means that \(\mathbf{z}\) does not belong to the \(k\)-th class (\(t_k \notin k\)-th class), and \(t_k=1\) means that \(\mathbf{z}\) belongs to the \(k\)-th class (\(t_k \in k\)-th class).  The conditional density \( f_{\boldsymbol{Z}}(\boldsymbol{z} | t_k = 0) \) is the density of \( \boldsymbol{z} \) given that it does not belong to the \( k \)-th Gaussian component. The detailed calculation of \( f_{\boldsymbol{Z}}(\boldsymbol{z} | t_k = 0) \) can be found in Appendix C. The probability that the data point \(\mathbf{z}\) belongs to the \(k\)-th class is denoted by \(f_{\boldsymbol{Z}}(t_k=1)\).  Consequently, the joint probability distribution of the data point \(\mathbf{z}\) and the class membership \(t_k\) can be expressed as 
\begin{equation}
    f_{\boldsymbol{Z}}(\mathbf{z},t_k=1)=f_{\boldsymbol{Z}}(\mathbf{z}|t_k=1)f_{\boldsymbol{Z}}(t_k =1)= \frac{e^{-\lambda}\lambda^k}{k!}\mathcal{N} \Big(\mathbf{z};\boldsymbol{\mu}_{k}^{(z)},\,{\boldsymbol{\Sigma}_{k}^{(z)}}\Big),
    \label{eq The joint probability}
\end{equation}
and the full generic model is given by 
\begin{equation}
f_{\boldsymbol{Z}}(\mathbf{z};w,\boldsymbol{\mu},\boldsymbol{\Sigma})=\sum_{t}^ {}f_{\boldsymbol{Z}}(\mathbf{z},t) =\sum_{k}\frac{e^{-\lambda}\lambda^{k}}{k!}\mathcal{N}\Big(\mathbf{z};\boldsymbol{\mu}_{k}^{(z)},\,{\boldsymbol{\Sigma}_{k}^{(z)}}\Big).
\end{equation}

\begin{figure*}
\centering
\includegraphics[width=0.9\textwidth]{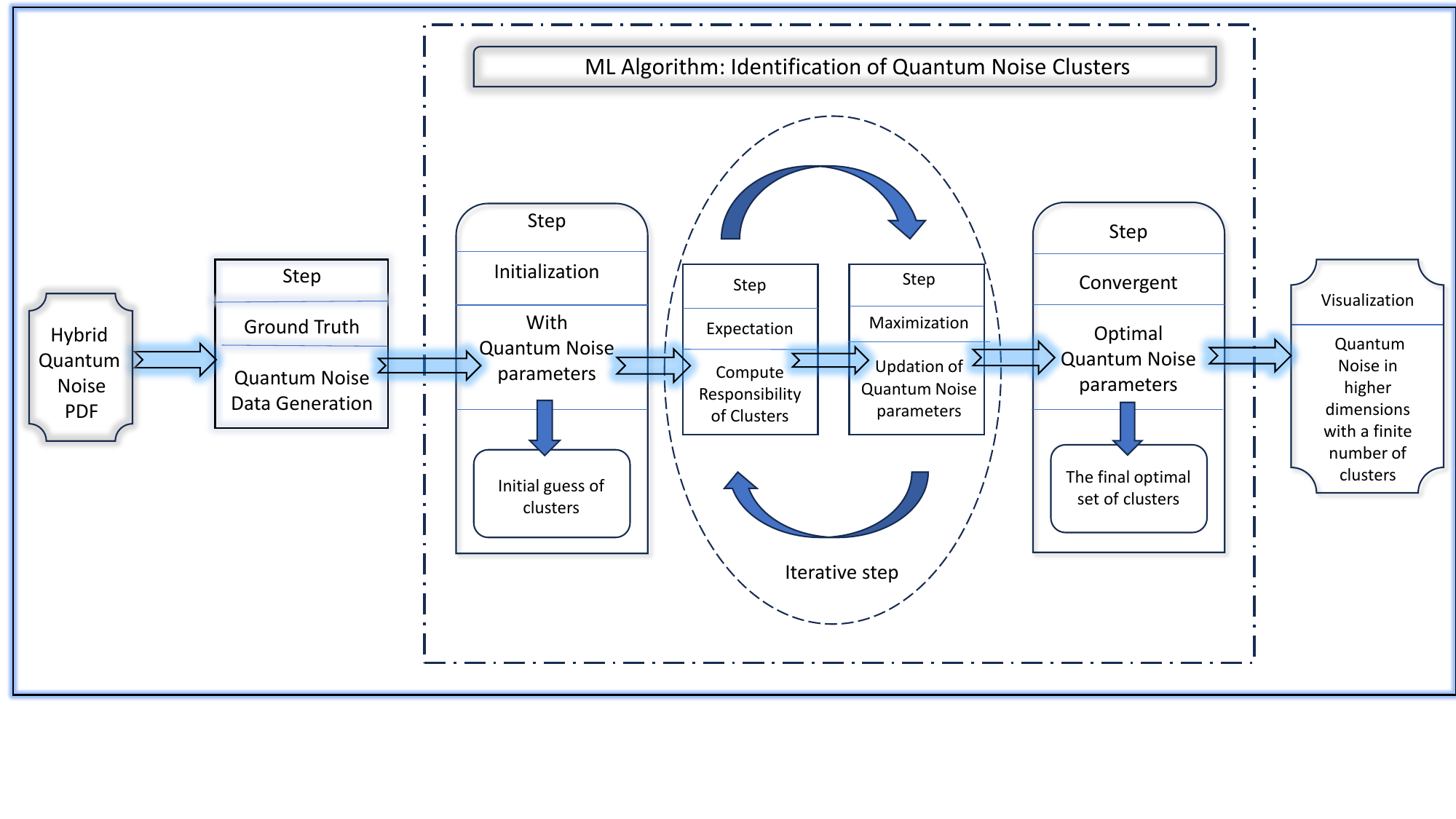} 
\caption{{The process of updating channel parameters and finding optimal clusters of hybrid quantum noise using \ac{ML} model : }Starts with the hybrid quantum noise model  \ac{p.d.f.} $\to$ ($1$) Step: Ground Truth as quantum hybrid noise data generation using the  \ac{p.d.f.}. The next step will feed this data into the  \ac{EM}algorithm $\to$  ($2$). Step: Initialize the \ac{EM}algorithm with estimated values for the hybrid quantum noise parameters, and in the next stages, it delivers iteratively optimized parameters during the E-step and M-step.  The algorithm's convergence gives the final optimized parametric values $\to$  ($3$). Step:  Visualization of clusters gives the initial estimated clusters, followed by the intermediate and final optimized clusters.}
\label{method}
\end{figure*}

The posterior or conditional probability of $t$ (the latent cluster) given a point $\mathbf{z}$ is 
\begin{multline}
f_{\boldsymbol{Z}}(t_k=1|\mathbf{z}) = \frac{f_{\boldsymbol{Z}}(t_k =1)f_{\boldsymbol{Z}}(\mathbf{z}|t_k=1)}{f_{\boldsymbol{Z}}(\mathbf{z})}\\=\frac{f_{\boldsymbol{Z}}(t_k =1)f_{\boldsymbol{Z}}(\mathbf{z}|t_k=1)}{\sum_{j}{}f_{\boldsymbol{Z}}(t_j =1)f_{\boldsymbol{Z}}(\mathbf{z}|t_j=1)}
= \frac{\frac{e^{-\lambda}\lambda^k}{k!}\mathcal{N} \Big(\mathbf{z};\boldsymbol{\mu}_{k}^{(z)},\,{\boldsymbol{\Sigma}_{k}^{(z)}}\Big)}{\sum_{j}{}\frac{e^{-\lambda}\lambda^j}{j!}\mathcal{N} \Big(\mathbf{z};\boldsymbol{\mu}_{k}^{(z)},\,{\boldsymbol{\Sigma}_{k}^{(z)}}\Big)}
=\gamma(z).
\end{multline}

After inferring a latent cluster for an observed point $\mathbf{z}$, probabilities can be assigned to this point that belongs to one of these unobserved clusters.  \(f_{\boldsymbol{Z}}(t_k=1|\mathbf{z})=\gamma(z)\) is the formula for obtaining the conditional posterior for the observed data point $\mathbf{z}$ that belongs to one of the latent clusters. We will call $\gamma(z)$ the responsibility that class $k$ takes to explain the data point $\mathbf{z}$.

The \ac{EM} algorithm produced a set of these parametric values $w, \boldsymbol{\mu}^{(z)}, \boldsymbol{\Sigma}^{(z)}$ to infer what parameters are the most optimal and which parameters lead to the distribution that most likely explains the hybrid quantum noise data set. Therefore, we need to define the log-likelihood function and want to maximize it to find optimal parameters that describe the hybrid quantum noise dataset. The log-likelihood of the hybrid quantum noise can be expressed as 
\begin{equation}
\begin{split}
	&\ln f_{\boldsymbol{Z}}\Big(\mathbf{z}; w, \boldsymbol{\mu}^{(z)},\,\boldsymbol{\Sigma}^{(z)}\Big)
	 =\sum_{n=1}^{N}\ln \Big(\sum_{k=1}^{K}\frac{e^{-\lambda}\lambda^k}{k!} \mathcal{N}(\mathbf{z}_{n} ; \boldsymbol{\mu}_{k}^{(z)}, \boldsymbol{\Sigma}_{k}^{(z)})\Big),
\end{split}    
\label{eq the log-likelihood}
\end{equation}
where $N$ is the total number of hybrid quantum noise data points named sample data. To maximize the log-likelihood function, in \eqref{eq the log-likelihood}, the expression cannot be further simplified due to the logarithmic summation over the total number of clusters. However, one of the solutions can be to use the \ac{EM} algorithm. One needs to improve the parameters and log-likelihood iteratively using this algorithm.

\subsection{Maximization of log-likelihood to find optimal parametric values for classifying the clusters: }

We need to maximize $ \ln f_{\boldsymbol{Z}}(\mathbf{z}; w,\boldsymbol{\mu}^{(z)}, \boldsymbol{\Sigma}^{(z)}) $ w.r.t. $\frac{e^{-\lambda}\lambda^k}{k!}, \boldsymbol{\mu}_{k}^{(z)}, \boldsymbol{\Sigma}_{k}^{(z)}$  $\forall k=1(1)K $, where $K$ is the total number of clusters. The problem is convex. Maximizing by looking for stationary points of logarithmic likelihood implies optimizing with respect to parameters $\frac{e^{-\lambda}\lambda^k}{k!}, \boldsymbol{\mu}_{k}^{(z)}, \boldsymbol{\Sigma}_{k}^{(z)}$. 
Therefore, we set the partial derivative of the log-likelihood function with respect to the mean vector \(\boldsymbol{\mu}_{k}^{(z)}\) of the \(k\)-th class to zero 
$\frac{\partial}{\partial \boldsymbol{\mu}_{k}^{(z)}} \ln f_{\boldsymbol{Z}}(\mathbf{z}; w,\boldsymbol{\mu}^{(z)}, \boldsymbol{\Sigma}^{(z)})=0$, 
and solve it for $\boldsymbol{\mu}_{k}^{(z)}$. However, this equation does not have a closed-form solution. The stationary points depend on the posterior $\gamma(t_{n,k})$. It can find local minima using the so-called alternative update algorithm of the (expected) posterior iterative algorithm $\gamma(t_{n,k})$ and parameters maximisation $\frac{e^{-\lambda}\lambda^k}{k!}, \boldsymbol{\mu}_{k}^{(z)}, \boldsymbol{\Sigma}_{k}^{(z)}$. 

If we take the derivative of log-likelihood, it depends on the parameters $\boldsymbol{\mu}_{k}^{(z)}, \boldsymbol{\Sigma}_{k}^{(z)}$ and $\frac{e^{-\lambda}\lambda^k}{k!}$. Instead of this, we could write the solution for $\boldsymbol{\mu}_{k}^{(z)}$ as a function of responsibility $\gamma(t_{n,k})$, that is,
\(\frac{\partial}{\partial \boldsymbol{\mu}_{k}^{(z)}} \ln f_{\boldsymbol{Z}}(z;\frac{e^{-\lambda}\lambda^k}{k!},\boldsymbol{\mu}_{k}^{(z)}, \mathbf{\Sigma_{k}^{(z)}})=0 \)
\(\implies \boldsymbol{\mu}_{k}^{(z)}\) is a function of \(\gamma(t_{n,k})\) and $\gamma(t_{n,k})$ depends on $\frac{e^{-\lambda}\lambda^k}{k!}, \boldsymbol{\mu}_{k}^{(z)}, \boldsymbol{\Sigma}_{k}^{(z)}$. The proposed solution is to group all remaining parameters in this term of responsibility. Therefore, the term of responsibility still depends on other parameters $ \frac{e^{-\lambda}\lambda^k}{k!}, \boldsymbol{\Sigma}_{k}^{(z)},$ even on $\mathbf{\mu_k}$. Using the \ac{EM} algorithm to fix the posterior probability or the responsibility $\gamma(t_{n,k})$ and solve for the case $\boldsymbol{\mu}_{k}^{(z)}$ and all the other parameters, one can update the responsibilities and again iterate this solution for maximization and solve for the parameters.

\begin{lem}
\label{lem:logLikelihoodmaximizationlem} The maximizing log-likelihood in \eqref{eq the log-likelihood} w.r.t. $\boldsymbol{\mu}_{k}^{(z)}$ 
by setting derivative of the log-likelihood w.r.t. $\boldsymbol{\mu}_{k}^{(z)}$ to $0$ that is
    $\frac{\partial}{\partial \boldsymbol{\mu}_{k}^{(z)}}\ln f_{\boldsymbol{Z}}(\mathbf{z} ; w,\boldsymbol{\mu}^{(z)},\mathbf{ \Sigma^{(z)}})=0$,
gives 
\begin{equation}
\boldsymbol{\mu}_{k}^{(z)}=\frac{\sum_{n=1}^{N}\gamma(t_{n,k})\mathbf{z}_n}{\sum_{n=1}^{N}\gamma(t_{n,k})}. 
\end{equation}
\end{lem}

\begin{IEEEproof}
Please refer to Appendix A.
\end{IEEEproof}

Now, to maximize the log-likelihood with respect to the mixing coefficient, $\ \frac {e^{-\lambda}\lambda^k}{k!}$ by considering a constrained optimization problem using the Lagrange multiplier method. To maximize log-likelihood w.r.t. $\frac{e^{-\lambda}\lambda^k}{k!}$ we have to set the derivative of the log-likelihood w.r.t. $\frac{e^{-\lambda}\lambda^k}{k!}$ to $0$, under the constraint that the sum of the probabilities $\sum_{k=1}^{K}\frac{e^{-\lambda}\lambda^k}{k!}$, add up to one, that is, $\sum_{k=1}^{K}\frac{e^{-\lambda}\lambda^k}{k!}=1$.

Let the Lagrange multipliers be denoted by
\begin{equation}
    L(\frac{e^{-\lambda}\lambda^k}{k!},\Lambda):=f(\mathbf{z}_n)+\Lambda(g(\mathbf{z}_n)-c),
\end{equation}
where $f(\mathbf{z}_n)= \sum_{n=1}^{N} \ln f_{\boldsymbol{Z}}(\mathbf{z}_n ;\frac{e^{-\lambda}\lambda^k}{k!}, \boldsymbol{\mu}_{k}^{(z)}, \boldsymbol{\Sigma}_{k}^{(z)})$ , $ g(\mathbf{z}_n)= \sum_{k=1}^{K}\frac{e^{-\lambda}\lambda^k}{k!}$ and $c=1$. Let us set the partial derivative of the Lagrange multipliers w.r.t. $\frac{e^{-\lambda}\lambda^k}{k!}$ that is $\frac{\partial}{\partial w_k}L(w,\Lambda)=0$ gives 
\begin{equation}
    \begin{split}
&\frac{\partial}{\partial\frac{e^{-\lambda}\lambda^{k}}{k!}}\Bigg[\sum_{n=1}^{N}\ln f_{\boldsymbol{Z}}(\mathbf{z}_n ;\frac{e^{-\lambda}\lambda^k}{k!}, \boldsymbol{\mu}_{k}^{(z)}, \boldsymbol{\Sigma}_{k}^{(z)})
+\Lambda\bigg\{\sum_{k=1}^{K}\frac{e^{-\lambda}\lambda^{k}}{k!}-1\bigg\}\Bigg]=0 \\
&\implies\sum_{n=1}^{N}\frac{\frac{\partial}{\partial\frac{e^{-\lambda}\lambda^{k}}{k!}}f_{\boldsymbol{Z}}(\mathbf{z}_n ;\frac{e^{-\lambda}\lambda^k}{k!}, \boldsymbol{\mu}_{k}^{(z)}, \boldsymbol{\Sigma}_{k}^{(z)})}{f_{\boldsymbol{Z}}(\mathbf{z}_n ;\frac{e^{-\lambda}\lambda^k}{k!}, \boldsymbol{\mu}_{k}^{(z)}, \boldsymbol{\Sigma}_{k}^{(z)})}
+\Lambda\frac{\partial}{\partial\frac{e^{-\lambda}\lambda^{k}}{k!}}\bigg\{\sum_{k=1}^{K}\frac{e^{-\lambda}\lambda^{k}}{k!}-1\bigg\}=0 \nonumber
\label{}
    \end{split}
\end{equation}

\begin{equation}
    \begin{split}
        &\implies\sum_{n=1}^{N}\frac{\frac{\partial}{\partial\frac{e^{-\lambda}\lambda^{k}}{k!}}\Big[\sum_{k=1}^{K}\frac{e^{-\lambda}\lambda^{k}}{k!}\mathcal{N}(\mathbf{z}_{n};\boldsymbol{\mu}_{k}^{(z)},\boldsymbol{\Sigma}_{k}^{(z)})\Big]}{\sum_{k=1}^{K}\frac{e^{-\lambda}\lambda^{k}}{k!}\mathcal{N}(\mathbf{z}_{n};\boldsymbol{\mu}_{k}^{(z)},\boldsymbol{\Sigma}_{k}^{(z)})} 
+\Lambda\frac{\partial}{\partial\frac{e^{-\lambda}\lambda^{k}}{k!}}\bigg\{\sum_{k=1}^{K}\frac{e^{-\lambda}\lambda^{k}}{k!}-1\bigg\}=0 \\
&\implies\frac{e^{-\lambda}\lambda^{k}}{k!}=-\frac{1}{\Lambda}\sum_{n=1}^{N}\gamma(t_{n,k}).
\label{}
    \end{split}
\end{equation}

Now we solve it for $\Lambda$. For that, we consider the constraint part \(\sum_{k=1}^{K}\frac{e^{-\lambda}\lambda^k}{k!}=1 .\).
Now, setting derivative of the Lagrange multipliers w.r.t. $\Lambda$ to zero, that is 
\begin{equation}
\begin{split}
   & \frac{\partial}{\partial \Lambda} L(\frac{e^{-\lambda}\lambda^k}{k!},\Lambda)=0  \implies \frac{\partial}{\partial \Lambda}\big\{f(\mathbf{z}_n)+\Lambda(g(\mathbf{z}_n)-c)\big\}=0 \\
   & \implies \sum_{k=1}^{K}\frac{e^{-\lambda}\lambda^k}{k!}-1 =0 
\implies  -\frac{1}{\Lambda}\sum_{n=1}^{N}\sum_{k=1}^{K}\gamma(t_{n,k})=1 
\implies \Lambda= -N,
\end{split}
\end{equation}
because  $\sum_{k=1}^{K}\gamma(t_{n,k})= \sum_{k=1}^{K} f_{\boldsymbol{Z}}(t_{n,k}|\mathbf{z})=1 $, this is due to the sum of all classes. Finally, this gives us 
\begin{equation}
w_{k}=\frac{1}{N}\sum_{n=1}^{N}\gamma(t_{n,k}). 
\end{equation}
Now we need to calculate the expression for $\boldsymbol{\Sigma}_{k}^{(z)}$.

\begin{lem}
\label{lem:LogLikelihoodMaximizationWRTSigma}  The maximization of the log-likelihood in \eqref{eq the log-likelihood} w.r.t. $\boldsymbol{\Sigma}_{k}^{(z)}$ by setting 
$\frac{\partial}{\partial \boldsymbol{\Sigma}_{k}^{(z)}} \ln f_{\boldsymbol{Z}}(\mathbf{z};w,\boldsymbol{\mu}, \boldsymbol{\Sigma})=0$,
gives
\begin{equation}
\boldsymbol{\Sigma}_{k}^{(z)} = \frac{\sum_{n=1}^{N} \gamma(t_{n,k}) (\mathbf{z}_n - \mu_k^{(z)})(\mathbf{z}_n - \mu_k^{(z)})^T}{\sum_{i=1}^{N} \gamma(t_{n,k})} .
\end{equation}
\end{lem}

\begin{IEEEproof}
Please refer to Appendix B. 
\end{IEEEproof} 
Now define the effective number of points in cluster K by 
$N_{k}= \sum_{n=1}^{N}\gamma(t_{n,k})$.
Hence, the solution for $\frac{e^{-\lambda}\lambda^k}{k!}, \mu_k^{(z)}, \boldsymbol{\Sigma}_{k}^{(z)}$ (dependent on the posterior) can be written as 
\begin{equation}
\begin{split}             
	\boldsymbol{\mu}_{k}^{(z)} = \frac{1}{N_k} \sum_{n=1}^{N} \gamma(t_{n,k}) \mathbf{z}_n,\, \, \,
	\frac{e^{-\lambda}\lambda^k}{k!} = \frac{N_k}{N}, \, \, \, 
	\boldsymbol{\Sigma}_{k}^{(z)} = \frac{1}{N_k} \sum_{n=1}^{N} \gamma(t_{n,k}) (\mathbf{z}_n - \boldsymbol{\mu}_{k}^{(z)})(\mathbf{z}_n - \boldsymbol{\mu}_{k}^{(z)})^T.
	\label{eq parameters}   
\end{split}
\end{equation} 
\subsection{Expectation-Maximization Algorithm (\ac{EM}):}

The \ac{EM} algorithm is a robust technique for estimating parameters in \acp{GMM}. This iterative algorithm comprises two primary steps: the expectation step (E-step) and the maximization step (M-step). Initially, the \ac{GMM} parameters are established, which involve initialization of the means \(\boldsymbol{\mu}_{k}^{(z)}\), covariances \(\boldsymbol{\Sigma}_{k}^{(z)}\), and the probabilities of the mixture component expressed as \(\frac{e^{-\lambda}\lambda^k}{k!}\) for each element \(k\). In the E-step, the algorithm updates the expected posterior probabilities, or responsibilities \(\gamma(z_{n,k})\), for each data point that concerns each component of the mixture. This step estimates the degree to which each component generates each data point. Subsequently, the M-step aims to maximize the log-likelihood of the model given the current parameter estimates, adjusting \(\boldsymbol{\mu}_{k}^{(z)}\), \(\boldsymbol{\Sigma}_{k}^{(z)}\), and the probabilities based on the fixed posteriors established in the E-step. The algorithm alternates between these E and M steps until it meets a convergence criterion, such as a negligible change in the log-likelihood or after a specified number of iterations. This iterative process allows the \ac{EM} algorithm to efficiently find local minima in the optimization landscape of the \ac{GMM} parameter estimation problem. In our probabilistic framework, we outline the \ac{EM} algorithm within Algorithm \ref{EMAlgo}.

\begin{algorithm}[t]

\caption{Expectation Maximization Algorithm for finding the optimal $\frac{e^{-\lambda}\lambda^k}{k!}, \boldsymbol{\mu}_{k}^{(z)}, \boldsymbol{\Sigma}_{k}^{(z)} $}

\label{EMAlgo}

\KwIn{ $\lambda$, $\boldsymbol{\mu}_{k,0}^{(z)}$, $\boldsymbol{\Sigma}_{k,0}^{(z)} $, $ k=1(1)K$}

\KwOut{ $\boldsymbol{\mu}_{k,m}^{(z)}$, $\boldsymbol{\Sigma}_{k,m}^{(z)} $}

$m\leftarrow1$

\Repeat{convergence }{
	
	\!\!\ Calculate the responsibilities $\gamma(z_{n,k,m})$ for each data point $\mathbf{z}_n$ and each component $k$ according to 
	$\gamma(z_{n,k}) = \frac{\frac{e^{-\lambda}\lambda^k}{k!} \mathcal{N}(\mathbf{z}_n;\boldsymbol{\mu}_{k}^{(z)}, \boldsymbol{\Sigma}_{k}^{(z)})}{\sum_{k=1}^{K} \frac{e^{-\lambda}\lambda^k}{k!} \mathcal{N}(\mathbf{z}_n;\boldsymbol{\mu}_{k}^{(z)}, \boldsymbol{\Sigma}_{k}^{(z)})}$\;

	\!\!\ $w_{k,m}=\frac{e^{-\lambda}\lambda^k}{k!} = \frac{1}{N} \sum_{n=1}^{N} \gamma(\mathbf{z}_{n,k,m}) $\;
	
	\!\!\ $\boldsymbol{\mu}_{k,m}^{(z)} = \frac{1}{N_k} \sum_{n=1}^{N} \gamma(\mathbf{z}_{n,k,m}) \mathbf{z}_n$ \;
	
	\!\!\ $\boldsymbol{\Sigma}_{k,m}^{(z)} = \frac{1}{N_k} \sum_{n=1}^{N} \gamma(\mathbf{z}_{n,k,m}) (\mathbf{z}_n - \boldsymbol{\mu}_{k}^{(z)})(\mathbf{z}_n - \boldsymbol{\mu}_{k,m}^{(z)})^T $\;
	
	$\!\!$$m\leftarrow m+1$\;
	
} \end{algorithm}

\begin{figure}[t]
    \centering
    \subfloat[\centering ]{{\includegraphics[width=7.5cm]{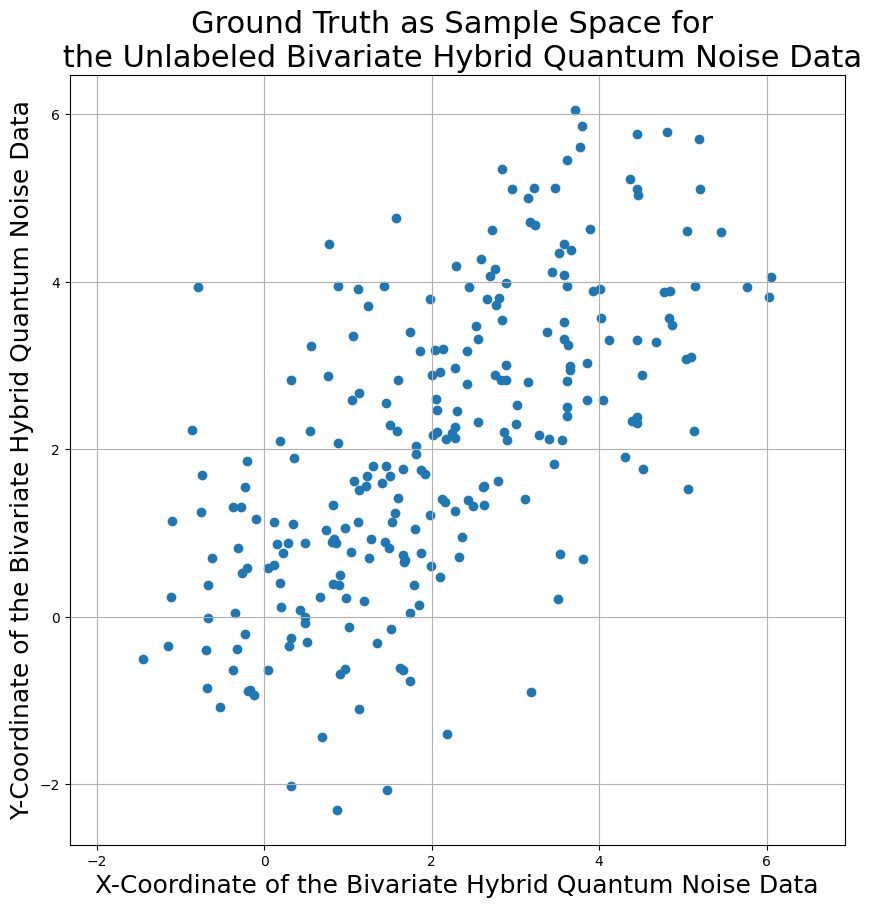} }}%
    \qquad
    \subfloat[\centering ]{{\includegraphics[width=7.5cm]{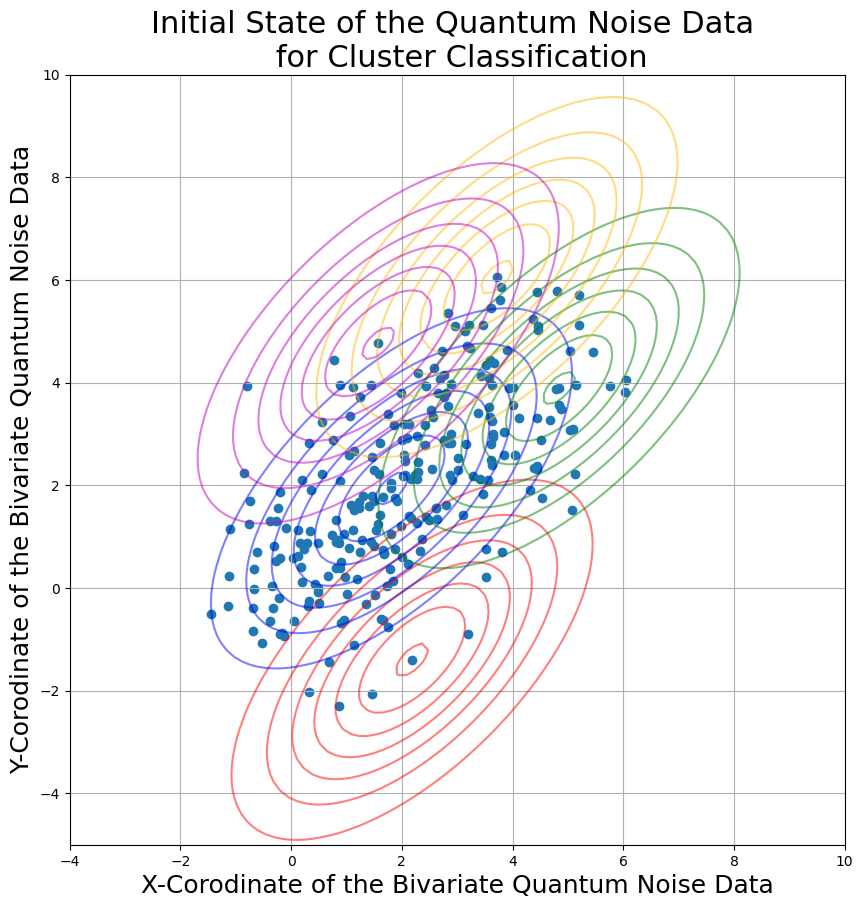} }}%
    \caption{(a) The ground truth data of bivariate quantum noise sampling generated as a convolution of additive white Gaussian noise and Poissonian quantum noise. In our studies, the ground truth of the hybrid quantum noise data has been generated from the component GMMs
    , (b) The initial guess of Gaussian mixture clusters of bivariate hybrid quantum noise sample data. Each of the five colored sets of spirals represents the hybrid quantum noise data points generated from that cluster, with each eye representing the mean of the respective Gaussian component.}%
    \label{fig ground truth and initial guess}%
\end{figure}



\subsection{The convergence of  \ac{EM} algorithm in the context of log-likelihood function}

Convergence of the \ac{EM} algorithm is typically understood in terms of the log-likelihood function of the observed data. In terms of monotonicity, the \ac{EM} algorithm ensures that the log-likelihood function of the observed data does not decrease with each iteration \cite{dempster1977maximumlikelihood}.
Under general conditions, the \ac{EM} algorithm converges to a local maximum (or saddle point in some cases) of the log-likelihood function. The convergence rate is typically linear but can slow near the maximum. The specific convergence rate can depend on the proximity of the initial parameter estimates to the actual values and the shape of the log-likelihood surface \cite{wu1983convergenceEM}. The convergence of the \ac{EM} algorithm is guaranteed under certain regularity conditions, including certain log-likelihood function derivatives and the parameter space's boundedness. However, these conditions can sometimes only be strictly met in practice, leading to convergence to local optima or, in rare cases, divergence if assumptions are violated severely \cite{mclachlan2007EMalgo}. For practical considerations, the \ac{EM} algorithm's convergence is usually monitored by checking the change in the log-likelihood value or the change in parameter estimates between successive iterations. 

\subsection{Discussion on selecting \ac{GMM} in comparison to existing \ac{ML} models}



 {In hybrid quantum noise modeling, \ac{GMM} excels due to its flexibility, probabilistic approach, and ability to handle complex noise distributions. It effectively models overlapping clusters and multiple sub-populations, such as those influenced by quantum shot noise and classical \ac{AWGN} \cite{Nielsen_Optical_cluster_states}. Unlike K-means \cite{patel2020clustering}, which provides hard clustering, \ac{GMM}'s soft clustering assigns probabilities to each data point, accommodating the overlapping distributions standard in quantum noise. An initial comparison with the K-Means algorithm was performed in this article; however, the study primarily focused on the \ac{GMM}-based \ac{EM} algorithm for clustering visualization. Techniques such as the Elbow Method, Silhouette Analysis, Gap Statistic, and probabilistic approaches like \ac{BIC} and \ac{AIC} are explored for evaluating clustering quality and determining the optimal number of clusters. The \ac{GMM} framework, combined with the \ac{EM} algorithm, was chosen for its iterative likelihood optimization, ability to handle missing data, and effectiveness in modeling overlapping clusters with soft probabilistic assignments. The \ac{EM} algorithm enhances \ac{GMM} by iteratively optimizing parameters like means and variances, ensuring robust modeling of noisy or incomplete quantum data. Its computational efficiency and quadratic time complexity make it suitable for real-time applications in quantum communication systems, such as satellite-based \ac{QKD} \cite{Mouli2024Asymp_QKD_SatComm, Mouli2024Finite_size_QKD_QComm}. Compared to advanced \ac{ML} techniques like \ac{DL} or \ac{RL} \cite{mackeprang2020rRL_quantum_clusters}, or Bayesian inference \cite{rath2020bayesianQuantumStates}, the \ac{EM} algorithm offers a practical balance between complexity, computational demands, and accuracy, making it ideal for hybrid quantum noise modeling.}

\begin{figure}[t]
    \centering
    \subfloat[\centering ]{{\includegraphics[width=7.5cm]{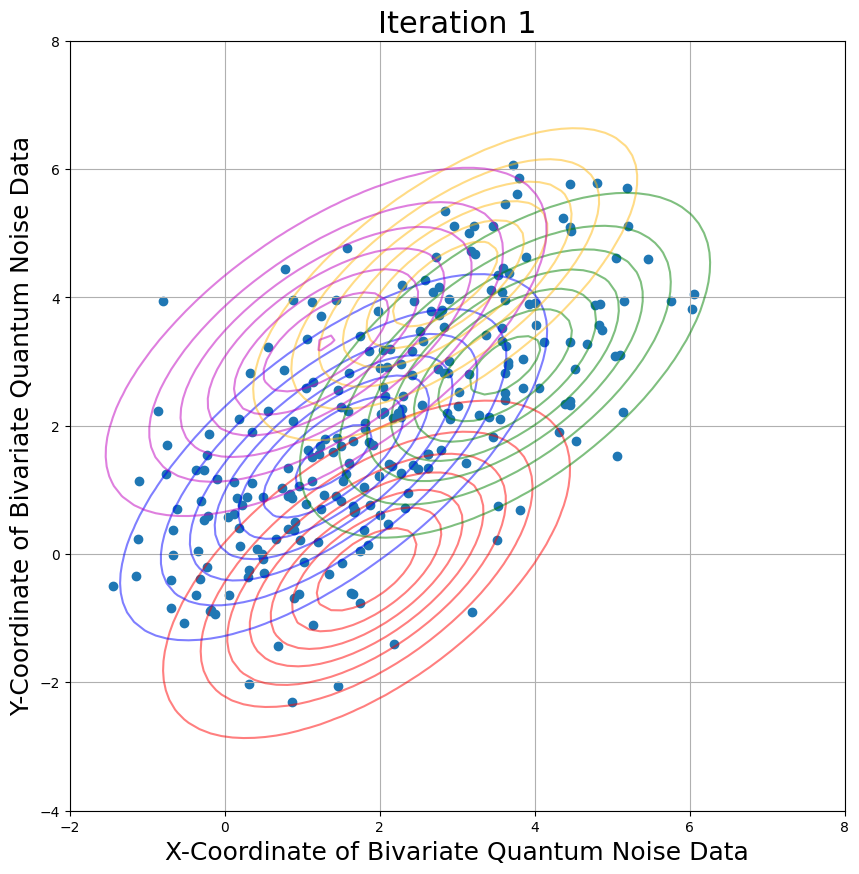} }}%
    \qquad
    \subfloat[\centering ]{{\includegraphics[width=7.5cm]{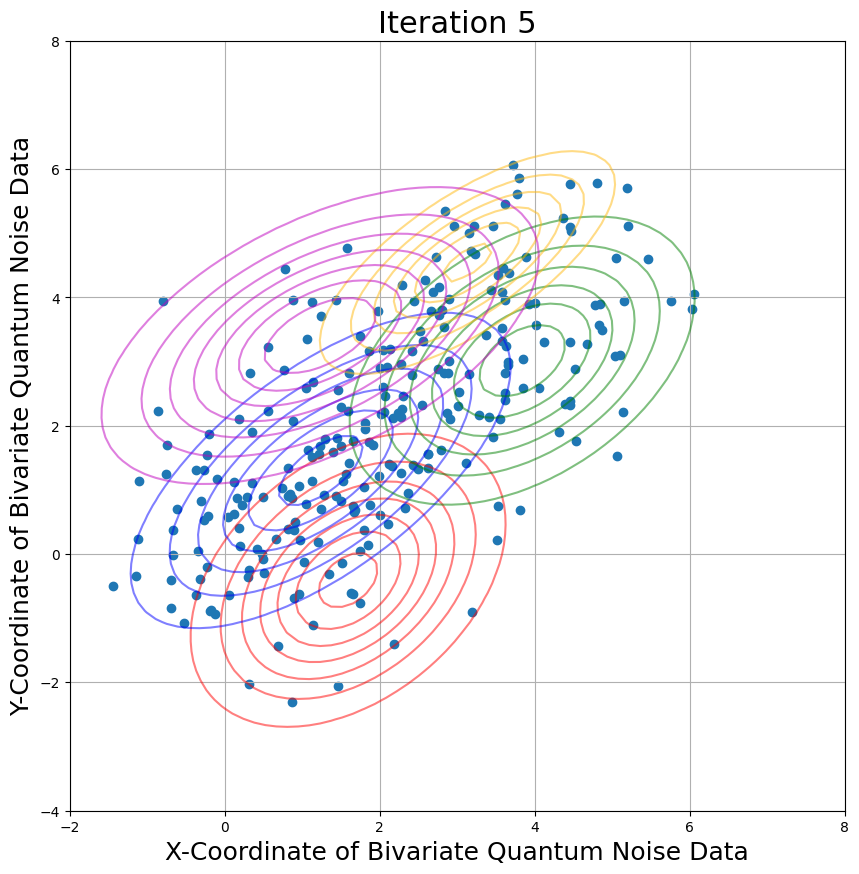} }}%
    \caption{(a) 1st iteration result under the updating of component of Gaussian mixture clusters of the bivariate hybrid quantum noise sample data. The change in the cluster's eyes and the locations results from the iteratively applying E-step and M-step in the  \ac{EM}algorithm
    , (b)The result after iteration size 5 for updating component of Gaussian mixture clusters of the bivariate hybrid quantum noise sample data.  }%
    \label{fig it1 and it5}%
\end{figure}



\section{Application in Maximization of Quantum Channel Capacity}

\vspace{0.15cm}

The interplay between the Poisson parameter and the \ac{GMM} is fundamental to accurately simulating quantum noise in quantum channel modeling. The Poisson parameter, $\lambda$, quantifies the average photon count detected over a specific interval, highlighting the quantum nature of light and the stochastic process of photon emission and detection. This parameter is essential for modeling photon statistics in quantum optics \cite{renker2009advances, Paul1982photon}, where photon arrivals at a detector are typically Poisson distributed, reflecting the quantized nature of light. On the contrary, the \ac{GMM} provides a probabilistic framework for modeling quantum noise, encompassing both quantum and classical noise influences on quantum states during transmission. This allows for a more comprehensive representation of various noise sources and their statistical behaviors in a quantum channel \cite{mouli2024, chakraborty2024hybridquantumnoiseapproximation, fox2006quantum}. The connection between these two models lies in their joint capability to more accurately represent the complexity of quantum noise. Although the Poisson parameter $\lambda$ captures the quantum mechanical aspects of the photon distribution, the \ac{GMM} accounts for a broader range of noise scenarios, including classical noise effects such as thermal noise and detector inefficiencies \cite{eisert2005gaussian}. This blend facilitates nuanced understanding and optimization of quantum channels, enabling the development of sophisticated error correction codes and communication protocols tailored to mitigate the specific noise characteristics encountered in quantum information transfer.

Taking into account hybrid quantum noise and the Gaussian transmitted signal model as discussed above \cite{mouli2024,chakraborty2024hybridquantumnoiseapproximation}, the capacity of the quantum channel can be expressed as follows: \\ 
\begin{equation}
 \begin{split}
   & C  =\sum_{i=0}^{R}\frac{e^{-\lambda}\lambda^i}{i!}\Bigg(-\log_{2}\Big(\frac{e^{-\lambda}\lambda^i}{i!}\Big)+ \frac{1}{2} \log_2{\Big((2\pi e)^{M}\big|{\boldsymbol{\Sigma}}_{i}^{(y)}\big|\Big)}\\
   & +\log_{2}\Big(\sum_{j=0}^{R}\frac{e^{-\lambda}\lambda^j}{j!} \mathcal{N} \Big(\boldsymbol{\mu}_{i}^{(z)};\boldsymbol{\mu}_{j}^{(z)},\,{\boldsymbol{\Sigma}}_{i}^{(z)}+{\boldsymbol{\Sigma}}_{j}^{(z)}\Big)\Big)\Bigg).
 \label{eq the expression for the channel capacity for Gaussian input vector}
 \end{split}
\end{equation}

This formula delineates the capacity of the quantum channel for scenarios where both the noise \(\boldsymbol{Z}\) and the received signal \(\boldsymbol{Y}\) are represented as random vectors of dimension size \(M\). In scalar analogy, that is, when the noise $Z$ has scalar \ac{p.d.f.}, 
the expression of capacity is reduced to
\begin{equation}
 \begin{split}
    C =\sum_{i=0}^{R}\frac{e^{-\lambda}\lambda^i}{i!}\Bigg[-\log_{2}\Big(\frac{e^{-\lambda}\lambda^i}{i!}\Big)+ \frac{1}{2}\log_2{\Big(2\pi e{\sigma}_{i}^{(y)}\Big)}\\
    +\log_{2}\Bigg(\sum_{j=0}^{R}\frac{e^{-\lambda}\lambda^j}{j!}\mathcal{N} \Big({\mu_{i}}^{(z)};{\mu_{j}}^{(z)},\,{{\sigma}_{i}^{(z)}}^2+{{\sigma}_{j}^{(z)}}^2\Big)\Bigg)\Bigg],
 \end{split}
  \label{eq the expression for the channel capacity Gaussian input scaler}
\end{equation}         
by putting $M=1$, and replacing $\big|{\boldsymbol{\Sigma}}_{i}^{(y)}\big|$ by ${\sigma}_{i}^{(y)}$, $\boldsymbol{\mu}_{i}^{(z)}$ by $\mu_{i}^{(z)}$, $\boldsymbol{\mu}_{j}^{(z)}$ by $\mu_{j}^{(z)}$, $ {\boldsymbol{\Sigma}}_{i}^{(z)}$ by ${{\sigma}_{i}^{(z)}}^2 $  and $\boldsymbol{\Sigma}_{j}^{(z)}$ by ${{\sigma}_{j}^{(z)}}^2$, where each vector is replaced by its scalar analogue \cite{mouli2024,chakraborty2024hybridquantumnoiseapproximation}. Again $\mu_{i}^{(z)}=\mu_{Z_2}+i$  $\forall i$, ${{\sigma}_{i}^{(z)}}^2={\sigma_{Z_2}}^2$ $\forall i$, and  ${{\sigma}_{i}^{(y)}}^2={\sigma_{X}}^2+{\sigma_{Z_2}}^2$  $\forall i$. Therefore, the capacity of the quantum channel is given by
\begin{equation}
\begin{split}
	C &
	=\sum_{i=0}^{R}\frac{e^{-\lambda}\lambda^i}{i!}\Bigg[-\log_{2}\Big(\frac{e^{-\lambda}\lambda^i}{i!}\Big)
	+ \frac{1}{2} \log_2{\Big(2\pi e \big({\sigma_{X}}^2+{\sigma_{Z_2}}^2\big)\Big)}\\
	& +\log_{2}\Bigg(\sum_{j=0}^{R}\frac{e^{-\lambda}\lambda^j}{j!}\frac{1}{\sqrt{2}\sigma_{Z_2}\sqrt{2\pi}}e^{-\frac{1}{2}\Big(\frac{i-j}{\sqrt{2}\sigma_{Z_2}}\Big)^2}\Bigg)\Bigg].
	\label{eq the final expression for the channel capacity Gaussian scaler input}
\end{split}
\end{equation}  


\begin{figure}[t]
    \centering
    \subfloat[\centering ]{{\includegraphics[width=7.5cm]{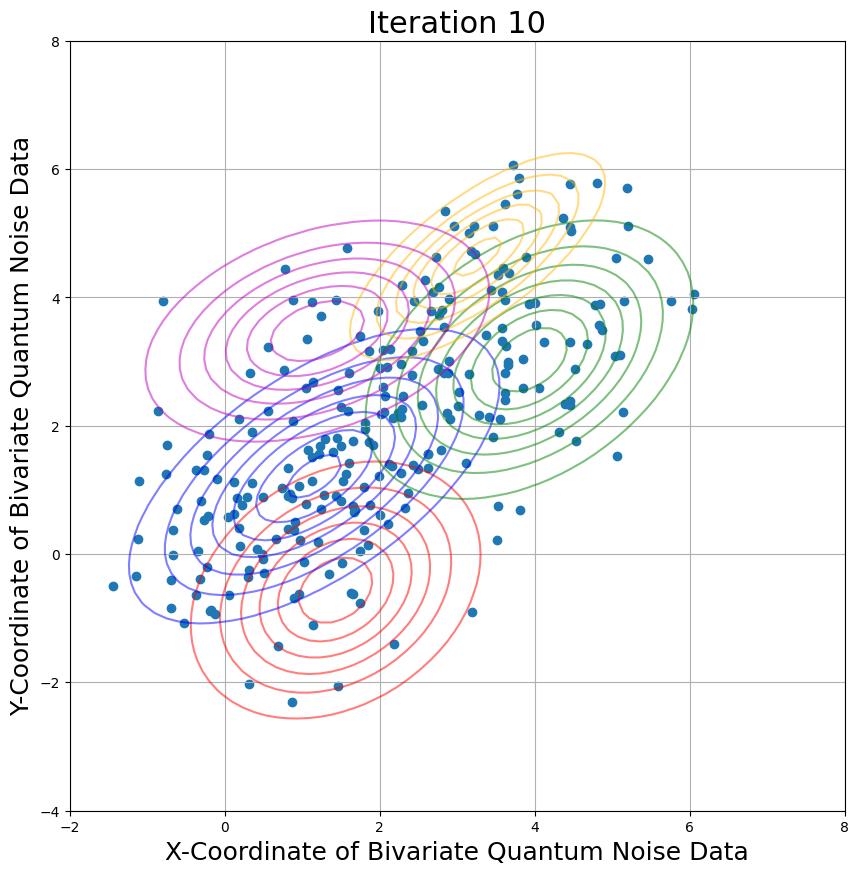} }}%
    \qquad
    \subfloat[\centering ]{{\includegraphics[width=7.5cm]{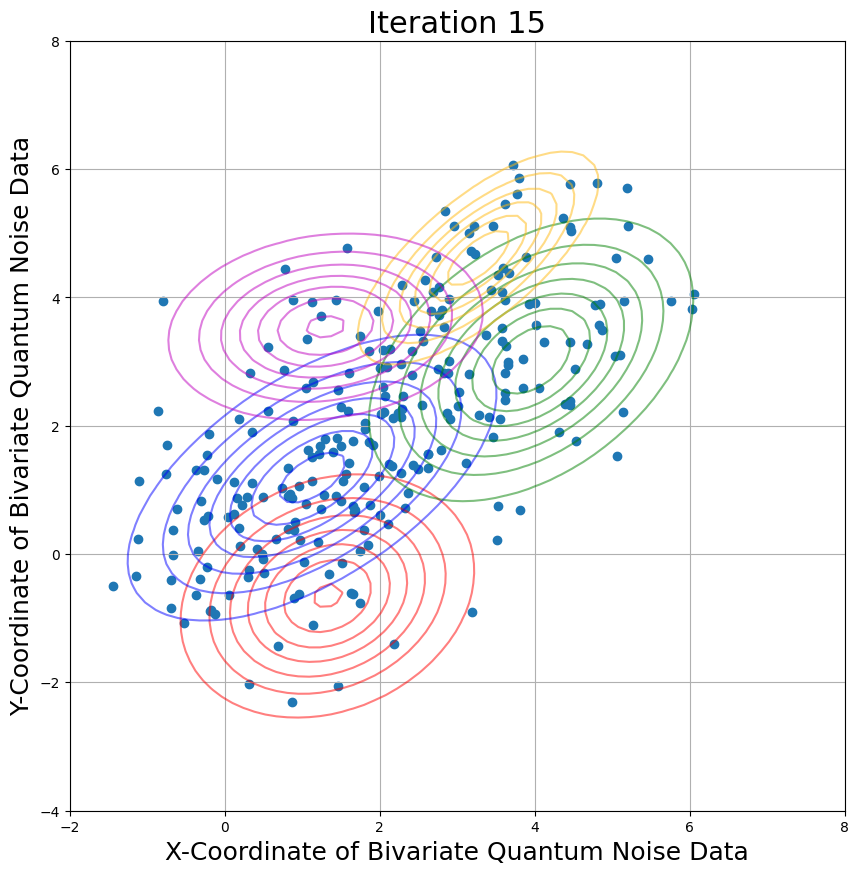} }}%
    \caption{(a) The result after iteration size 10 for updating component of Gaussian mixture clusters of the bivariate hybrid quantum noise sample data
    , (b) The result after iteration size 15 under the updating of component of Gaussian mixture clusters of the bivariate hybrid quantum noise sample data.}%
    \label{fig it10 and it15}%
\end{figure}



\section{Results and Numerical Analysis}

In this numerical section, we have visualized the implementation of the model. In the first part, we discussed how to visualize hybrid quantum noise in terms of Gaussian mixture clusters with weightage from Poisson distribution and updated the parameters associated with the channel's capacity. In the second part, we investigated how the capacity changes with updated parametric values with respect to the initial guess of parameters, and this leads to the final contribution of this work in terms of the better achievable capacity of the quantum channel. 

\begin{figure}[t]
    \centering
    \subfloat[\centering ]{{\includegraphics[width=7.5cm]{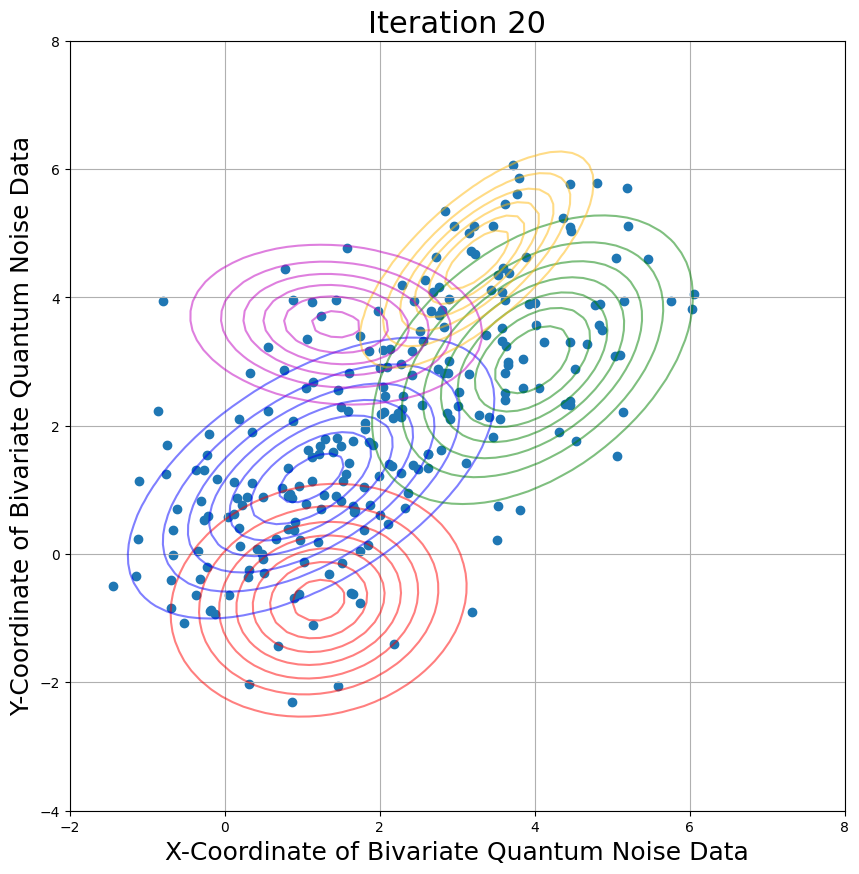} }}%
    \qquad
    \subfloat[\centering ]{{\includegraphics[width=7.5cm]{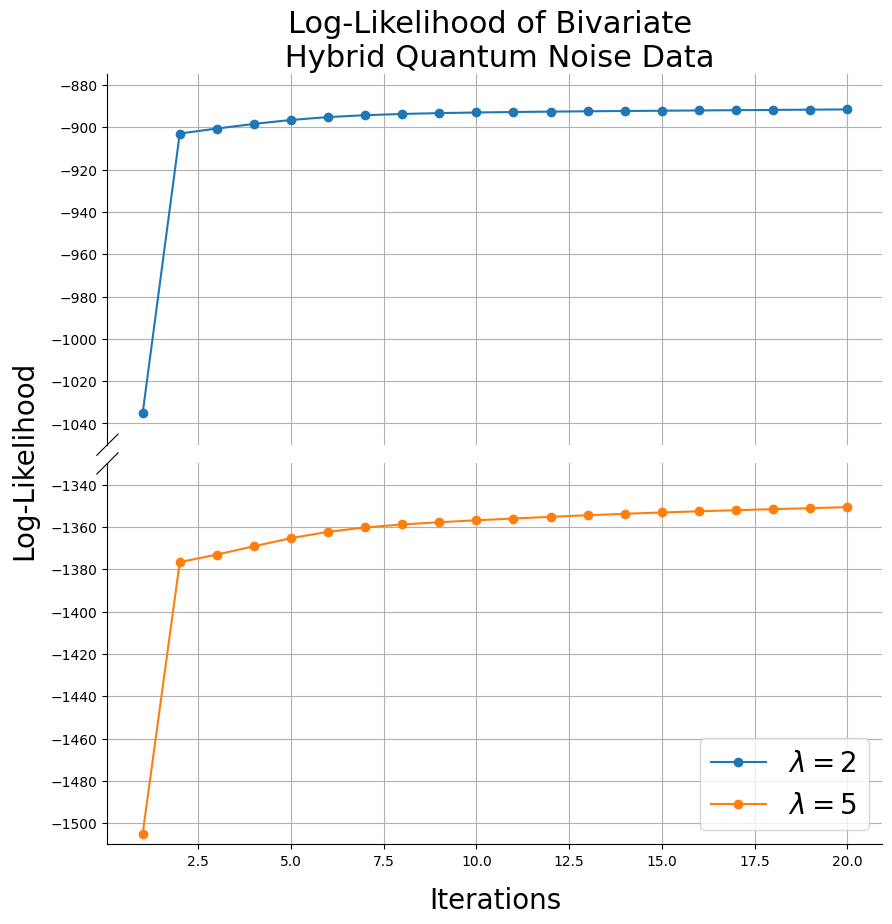} }}%
    \caption{(a) The Final results corresponding to iteration size 20 under the updating of component of Gaussian mixture clusters of the bivariate hybrid quantum noise sample data
    , (b) The convergence of log-likelihood of the bivariate hybrid quantum noise probability density function for $\lambda=2$ and $5$. Curve saturation implies the convergence of the \ac{EM}algorithm for cluster finding.}%
    \label{fig Final it20 and log_likelihood_2_5}%
\end{figure}

Note that the Section ~\ref{System Model} ~\ref{Gaussian Quantum Channel equation}, it considered the dimension reduction in the qubit. This representation leads $1$-D qubit and the corresponding scalar \ac{p.d.f.}. However, to visualize the sample from vectorized \ac{p.d.f.}, we consider the bivariate quantum-noise data sampled from \eqref{eq approx pdf of joint quantum noise in Gaussian form for vector} shown in Fig.~\ref{fig ground truth and initial guess} to Fig.~\ref{fig Final it20 and log_likelihood_2_5}. The bivariate hybrid quantum noise data sample has been plotted where the X-axis and Y-axis of the plots represent the X-coordinate and Y-coordinate of the bivariate qubit vectors.

\subsection{The error tolerance in hybrid quantum-noise data visualization}

 {In this subsection, we visualize the hybrid quantum noise dataset in the vector field and investigate its properties based on the hybrid quantum noise model defined by \eqref{eq approx pdf of joint quantum noise in Gaussian form for vector}. The model uses a \ac{GMM} consisting of Gaussian components weighted by coefficients derived from a Poisson distribution, where the weights represent the prior probabilities of cluster membership. The weights are determined by \( f_{\boldsymbol{Z}}(t_k = 1) = \frac{e^{-\lambda}\lambda^k}{k!} \), ensuring that \( \sum_{k=1}^{K}\frac{e^{-\lambda}\lambda^k}{k!} \approx 1 \) for a chosen accuracy level, typically \( > 85\% \). The parameter \( \lambda \) controls the number of components \( K \), with higher \( \lambda \) leading to more components. For simplicity in visualization, a smaller \( \lambda \) is preferred. For instance, with \( \lambda = 2 \), the weights \( 0.271, 0.271, 0.135, 0.180, 0.009 \) sum to \( 0.947 \), allowing \( K = 5 \) clusters. Similarly, with \( \lambda = 5 \), the weights \( 0.175, 0.175, 0.146, 0.140, 0.104, 0.084, 0.065 \) sum to \( 0.889 \), resulting in \( K = 7 \) clusters. However, using only the five highest weights from \( \lambda = 5 \) yields a sum \( < 0.85 \), failing the initial condition. This illustrates how \( \lambda \) and the chosen accuracy influences the number of clusters. 
}

 {Higher values of \( \lambda \) produce a flatter Poisson distribution and lower mixing coefficients (\( w_i \)), leading to more components in the Gaussian mixture. To simplify the visualization of classified clusters in modeling quantum data, a lower \( \lambda \) is preferred. To ensure the constraint \( \sum_{i=0}^{R} w_i \approx 1 \) is satisfied with an error tolerance of $15\%$, we set an accuracy level of at least $85\%$ (\( |1 - \sum_{i=0}^{R} w_i| \leq 0.15 \)).
In quantum cryptography, Alice's light source must emit single photons to minimize eavesdropping risks. A low mean photon number (\( \lambda \)) is achieved by attenuating a pulsed laser, ensuring most pulses are photon-free or contain a single photon. For example, \( \lambda = 0.1 \) is common, but for this study, \( \lambda = 2 \) is used to focus on hybrid noise clustering and channel capacity.
To determine the optimal number of clusters (\( K \)) with at least $85\%$ accuracy, the most probable mixing coefficients for \( \lambda = 2 \) are chosen: \( w_0 = 0.135 \), \( w_1 = 0.271 \), \( w_2 = 0.271 \), \( w_3 = 0.18 \), and \( w_4 = 0.09 \). The sum \( \sum_{i=0}^{4} w_i = 0.947 \), yield an accuracy of $95\%$ and an error of $5\%$, within the tolerance limit. Thus, the optimal number of clusters is \( K = 5 \), satisfying the model's requirements.}

\subsection{The optimal number of cluster identifications using different Machine learning evaluation metrics}

\begin{figure}[t]
    \centering
    \subfloat[\centering ]{{\includegraphics[width=9.5cm]{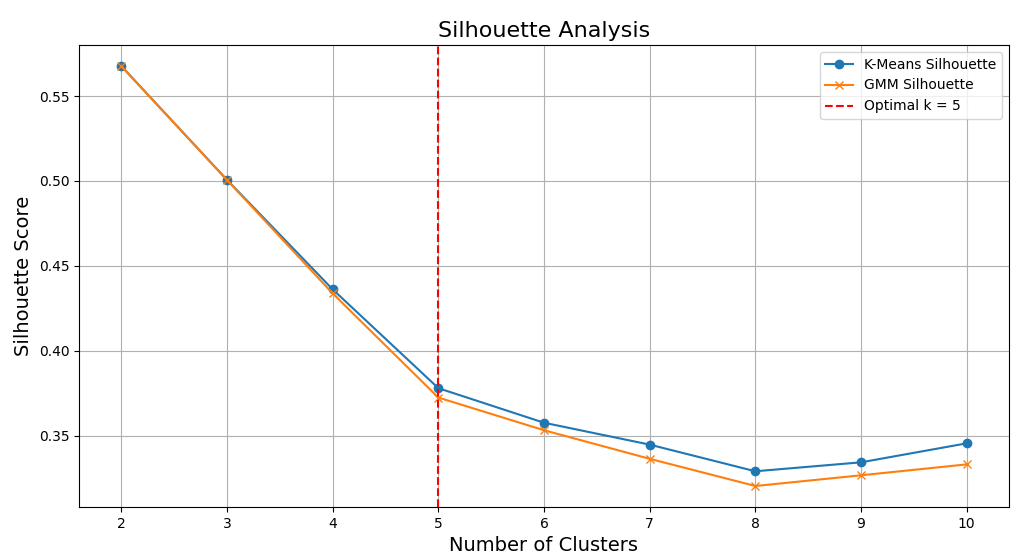} }}%
    \qquad
    \subfloat[\centering ]{{\includegraphics[width=9.5cm]{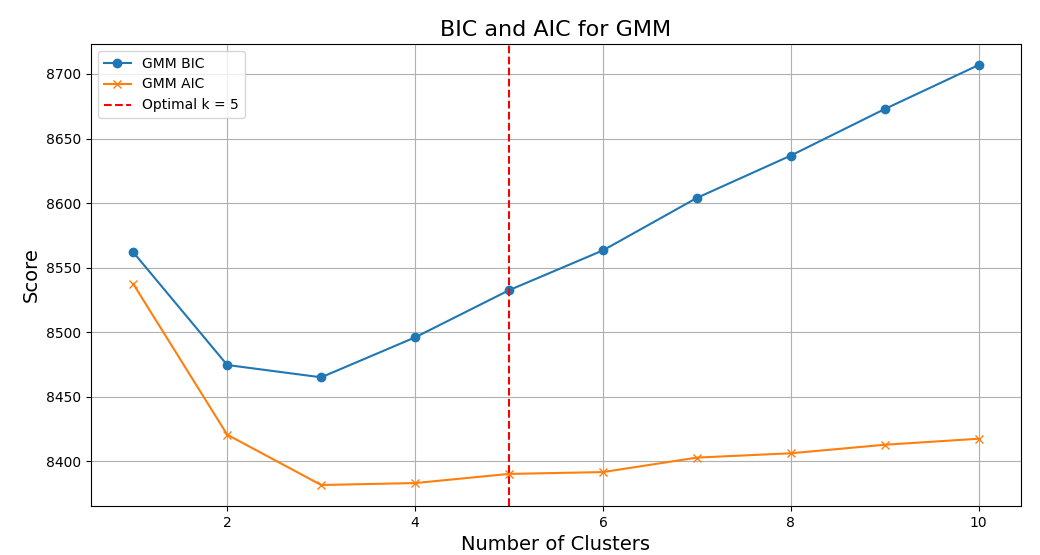} }}%
    \caption{(a) Silhouette Analysis evaluates cluster quality by measuring intra-cluster cohesion and inter-cluster separation, identifying \( k = 5 \) as the optimal cluster count with the highest average silhouette score for the hybrid quantum noise data, (b) Bayesian Information Criterion (BIC) and Akaike Information Criterion (AIC) analyses confirm \( k = 5 \) as the optimal cluster count, minimizing overfitting while capturing the underlying structure of the hybrid quantum noise data.
    }%
    \label{fig Silhouette Analysis and BIC and AIC}%
\end{figure}

 {Determining the optimal number of clusters (\( k \)) is a critical step in unsupervised learning, as it directly impacts the performance and interpretability of clustering algorithms. This study began with 10 initial clusters generated to represent the complexity of hybrid quantum noise data. Through a detailed analysis incorporating several established methods, it was determined that \( k = 5 \) was the optimal number of clusters, effectively capturing the underlying data structure while avoiding overfitting.
Among the methods used, Silhouette Analysis proved particularly effective for evaluating cluster quality by measuring how well separated each data point was from other clusters. For a given data point \( i \), the silhouette coefficient \( s(i) \) is calculated as \( s(i) = \frac{b(i) - a(i)}{\max(a(i), b(i))} \), where \( a(i) \) is the average distance to points within the same cluster (intra-cluster distance) and \( b(i) \) is the average distance to points in the nearest cluster (inter-cluster distance). Silhouette scores range from -1 to 1, with scores near 1 indicating well-separated clusters, scores near 0 suggesting overlapping clusters, and negative scores implying incorrect clustering. The average silhouette score across all points provided a robust global measure, and the highest score corresponded to \( k = 5 \), indicating optimal separation and cohesion for this dataset.
In addition to Silhouette Analysis, the Elbow Method, Gap Statistic, and probabilistic approaches such as \ac{BIC} and  \ac{AIC} were employed to validate the results shown in Fig.~\ref{fig Silhouette Analysis and BIC and AIC}. The Elbow Method was identified \( k = 5 \) as the point where the rate of decrease in the \ac{WCSS} slowed significantly, balancing compactness and overfitting. The Gap Statistic confirmed this result by comparing the clustering structure of the observed data to a null reference distribution. For the \ac{GMM} used in this study, both \ac{BIC} and \ac{AIC} values were minimized at \( k = 5 \) shown in Fig.~\ref{fig Silhouette Analysis and BIC and AIC} (b). 
Starting with $10$ initial clusters allowed for an exploratory phase to assess the data's complexity. By systematically applying these methods, the reduction to \( k = 5 \) revealed the most meaningful clustering structure. This result highlights the importance of combining complementary metrics like Silhouette Analysis, Elbow Method, Gap Statistic, and probabilistic criteria to ensure robust and reliable clustering outcomes, particularly in complex datasets like hybrid quantum noise.}

 {We reformulate our noise probability density function using a \ac{GMM} with five components:
\(
f_{\boldsymbol{Z}}(\mathbf{z}) = \sum_{k=0}^{4} w_{k} \cdot \mathcal{N} \big(\mathbf{z}; \boldsymbol{\mu}_{k}^{(z)}, \boldsymbol{\Sigma}_{k}^{(z)} \big).
\)
Our goal is to determine the optimal parameters \( w_k \), \( \boldsymbol{\mu}_k^{(z)} \), and \( \boldsymbol{\Sigma}_k^{(z)} \) for \( k = 0, 1, 2, 3, 4 \). While a typical \ac{GMM} with five components requires 15 parameters (five each for weights, means, and covariances), we demonstrate that fewer parameters are sufficient for this representation.
The simulation, conducted in an Anaconda Jupyter Notebook with Python, employed the \ac{EM} algorithm to optimize the parameters and classify clusters in a hybrid quantum-classical dataset. The initial parameters were defined as follows:
\begin{equation}
    \begin{split}
      & \text{Weights: Poisson-distributed as } \,  w_0 = 0.135, w_1 = 0.271, w_2 = 0.271, w_3 = 0.18, w_4 = 0.09.\\
      & \text{Means:} \, \boldsymbol{\mu}_k^{(z)} = (\mu_{Z_2} + k) \mathbf{I}_{2 \times 1}, \,\text{where} \, \mu_{Z_2} = 0, \, k=0,1,2,3,4.\\
& \text{Covariances:} \,  \boldsymbol{\Sigma}_k^{(z)} = \sigma_{Z_2}^2 \mathbf{I}_{2 \times 2}, \, \text{where} \, \sigma_{Z_2}^2 = 1,  \, k=0,1,2,3,4.\\
 \nonumber
    \end{split}
\end{equation}
Figures illustrate the dataset and intermediate cluster estimations. For instance, Fig.~\ref{fig ground truth and initial guess} visualize the initial quantum noise distribution, while subsequent figures depict the evolution of clusters during iterations. Convergence is confirmed in Fig.~\ref{fig Final it20 and log_likelihood_2_5}, where the logarithmic likelihood stabilizes, affirming optimal parameter values.
The final parameter values after 20 iterations are:\\
\begin{equation}
    \begin{split}
       & \, w_0' = 0.1871, w_1' = 0.3321, w_2' = 0.2916, w_3' = 0.1487, w_4' = 0.0403 ,\\
& \, \boldsymbol{\mu}_0' = \begin{bmatrix} 0.4180 \\ 0.0895 \end{bmatrix}, 
\, \boldsymbol{\Sigma}_0' = \begin{bmatrix} 0.8910 & -0.1011 \\ -0.1011 & 0.9452 \end{bmatrix}.\\ \nonumber
    \end{split}
\end{equation}
These results validate the algorithm's ability to optimize Gaussian components while preserving the Poissonian parameter \(\lambda \approx 2\), corresponding to photon counts in the system. The updated parameters will be instrumental in analyzing the quantum channel capacity for Gaussian signals.}

Therefore, as a verification, this application of the \ac{EM} algorithm for finding the optimal cluster in the classification of bivariate hybrid-quantum data can update the optimal parametric values of Gaussian components and helps to identify the optimal clusters but does not affect the Poissonian parameter $\lambda$, which is the number of photon counts in the system. However, the iterative \ac{EM} algorithm has successfully updated the other two channel parameters. We will use the updated parameters to compare the quantum channel capacity for the Gaussian transmitted signal.


\begin{figure}[t]
    \centering
    \subfloat[\centering ]{{\includegraphics[width=7.5cm]{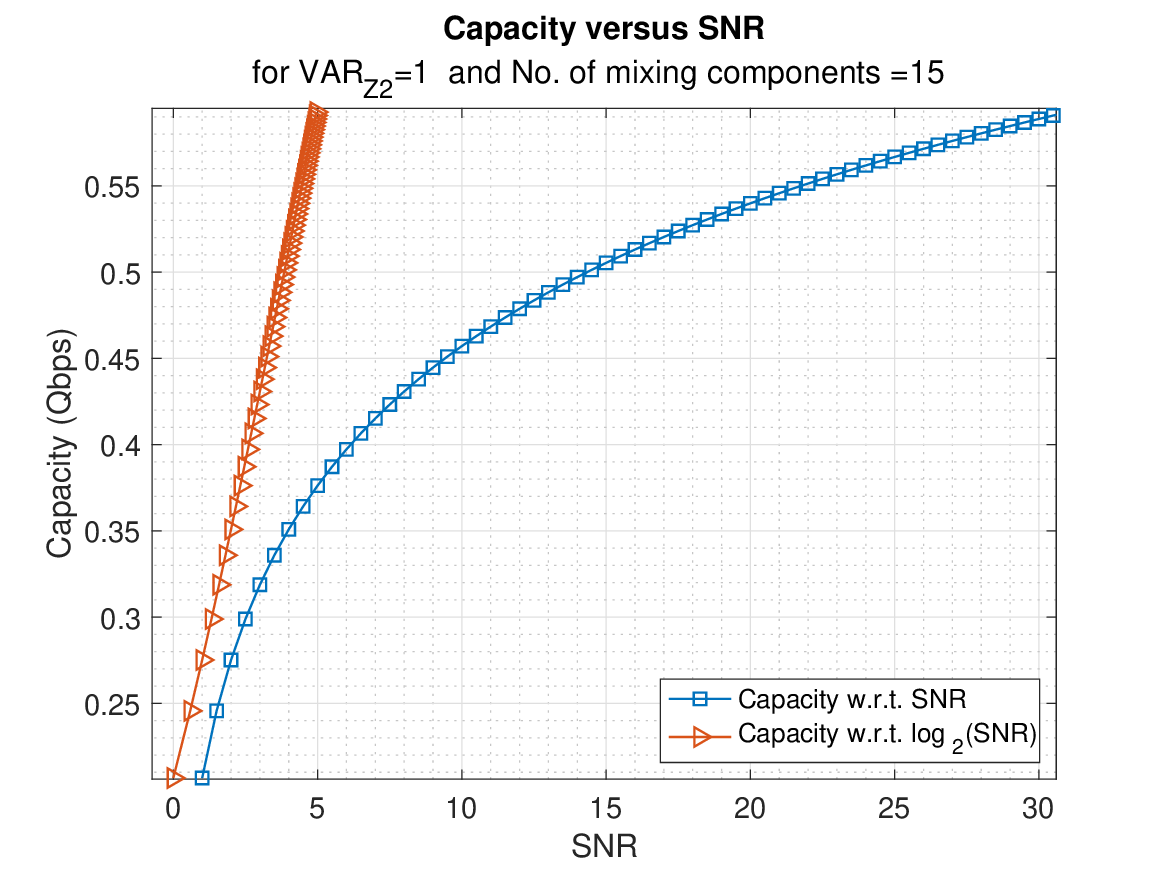} }}%
    \qquad
    \subfloat[\centering ]{{\includegraphics[width=7.5cm]{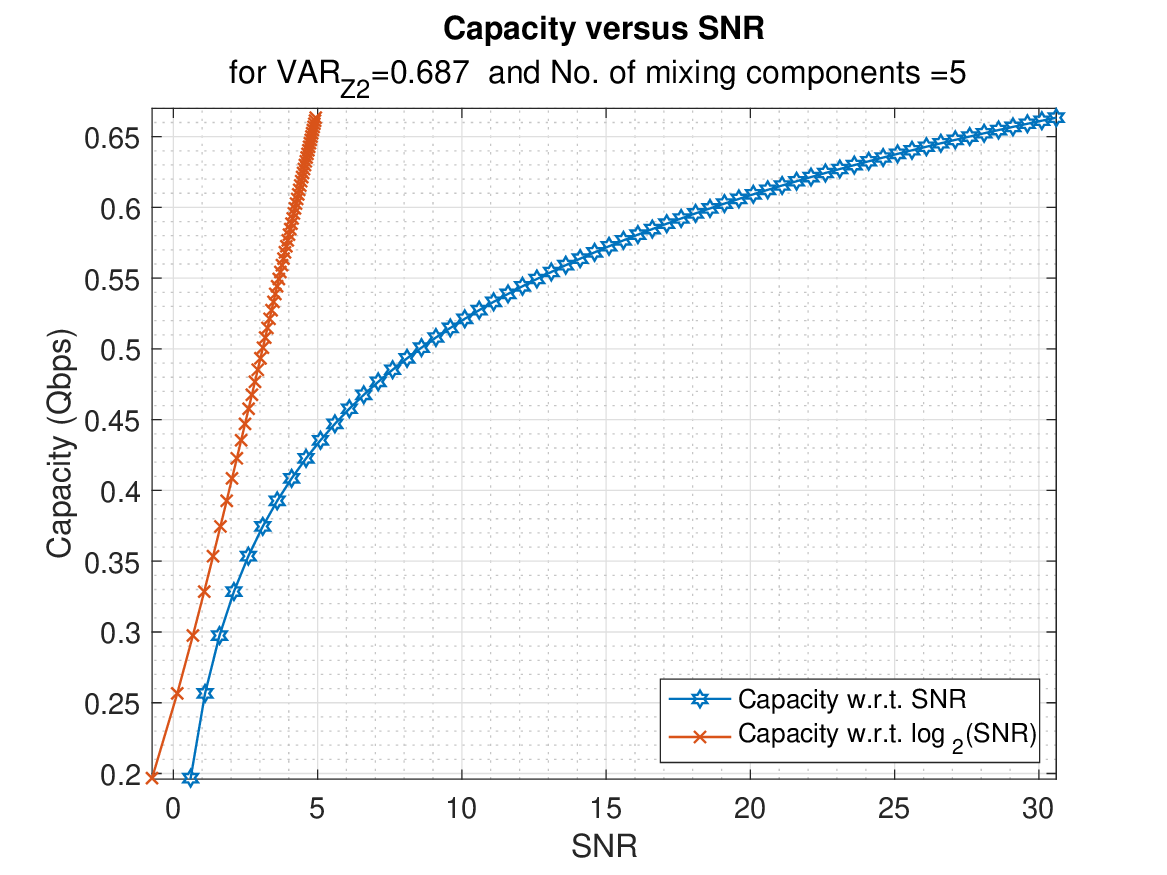} }}%
    \caption{(a) The capacity of the quantum channel with respect to \ac{SNR} for the number of components $15$ and variance $\sigma_{Z_2}^2 =1$
    , (b) The capacity of the quantum channel with respect to \ac{SNR} for the number of components $5$ and variance $\sigma_{Z_2}^2=0.687$. }%
    \label{fig capSNRforsigmaN2_initial and capSNRforsigmaN2_updated}%
\end{figure}



\subsection{Capacity comparison based on updated channel parameters and initial guess of parameters}

Consider 
the expression of quantum channel capacity in \eqref{eq the final expression for the channel capacity Gaussian scaler input} involving the channel parameter $\sigma_{Z_2}$ only to reflect any change in capacity, as we have shown that there is no change in the value of $\lambda$, and $\mu_{Z_{2}}$ does not involve the expression. So, start with the initial value of $\boldsymbol{\Sigma}_0= \sigma_{Z_{2}}^2\mathbf{I}_{2 \times 2}$
where 
$\sigma_{Z_{2}}=1$  $\longleftrightarrow$ $|\boldsymbol{\Sigma}_0|=1$,
and updated value of $\boldsymbol{\Sigma}_0'$
$\implies |\boldsymbol{\Sigma}_0'|=0.829 \longleftrightarrow \sigma_{Z_{2}}=0.829 \implies \sigma_{Z_{2}}^2=0.687$.  

This implies that the \ac{EM} algorithm will lower the Gaussian noise variance by optimal clustering. Now, let us compare the capacity with respect to the \ac{SNR} {for the following case studies: (I) $\lambda =2$, initial guess of $\sigma_{Z_{2}}^2=1$, and cluster size $15$; and (II) $\lambda =2$, the updated $\sigma_{Z_{2}}^2=0.687$, and optimized the cluster size $5$.} In Fig.~\ref{fig capSNRforsigmaN2_initial and capSNRforsigmaN2_updated} (a), we have plotted the benchmark result as the capacity of the quantum channel with respect to \ac{SNR} for the number of components {$15$} and the variance of Gaussian noise, $\sigma_{Z_{2}}^2=1$. Fig.~\ref{fig capSNRforsigmaN2_initial and capSNRforsigmaN2_updated} (b), we plot the capacity of the quantum channel with respect to the \ac{SNR} for the number of components $5$ and the variance $\sigma_{Z_{2}}^2 =0.687$. ML-optimized parameters  gives a better capacity estimation result than the initial estimated parameters.

\section{Conclusion}

{This research focuses on optimising hybrid quantum noise clustering to enhance the capacity of Gaussian quantum channels through advanced \ac{ML} techniques. Specifically, the study utilizes \ac{GMM} and the \ac{EM} algorithm to capture the intricate noise characteristics of  quantum channels. Hybrid quantum noise, consisting of quantum shot noise and classical \ac{AWGN}, is represented as an infinite mixture of Gaussian distributions weighted by Poissonian parameters. A key innovation of this work lies in reducing the number of clusters in the noise model and streamlining visualization while maintaining accuracy in channel capacity estimation without sacrificing the core properties of the noise. Notable contributions include minimizing the number of Gaussian clusters within acceptable error tolerances and applying the \ac{EM}-algorithm  to update quantum channel parameters, which results in improved channel capacity. The proposed method is validated through simulations, revealing that \ac{ML} -enhanced quantum noise clustering  significantly boosts performance in satellite-based quantum communication systems, particularly in \ac{QKD} applications. This study demonstrates that \ac{GMM} and \ac{EM}-algorithm offer a practical and efficient solution for modeling complex quantum noise in real-time applications, thereby advancing the optimization of quantum communication networks.}

\section*{APPENDIX}

\subsection{\label{proof:ofloglikelihoodmaximization}Proof of Lemma 1} 

The aim is to compute an expression for $\boldsymbol{\mu}_{k}^{(z)}$ as a function of responsibility $\gamma$; this can be done by setting a partial derivative of log-likelihood w.r.t. $\boldsymbol{\mu}_{k}^{(z)}$ as zero, given  by 
\begin{multline}
\frac{\partial}{\partial \boldsymbol{\mu}_{k}^{(z)}}\ln f_{\boldsymbol{Z}}(\mathbf{z} ; w,\boldsymbol{\mu}^{(z)},\mathbf{ \Sigma^{(z)}})=0
	\implies \frac{\partial}{\partial \boldsymbol{\mu}_{k}^{(z)}}\sum_{n=1}^{N}\ln f_{\boldsymbol{Z}}(\mathbf{z}_{n} ; w,\boldsymbol{\mu}^{(z)}, \boldsymbol{\Sigma}^{(z)})=0\\     
	\implies \sum_{n=1}^{N}\frac{\frac{e^{-\lambda}\lambda^k}{k!}\frac{\partial}{\partial \mu_k^{(z)}}\{\mathcal{N}\big(\mathbf{z}_{n} ; \boldsymbol{\mu}_{k}^{(z)}, \boldsymbol{\Sigma}_{k}^{(z)}\big)\}}{\sum_{k=1}^{K} \frac{e^{-\lambda}\lambda^k}{k!} \mathcal{N}(\mathbf{z}_{n} ; \boldsymbol{\mu}_{k}^{(z)}, \boldsymbol{\Sigma}_{k}^{(z)})} =0 
	\label{eq maximize w.r.t. mu 1}   
\end{multline}
To evaluate the density derivative of $\mathcal{N}\big(\mathbf{z}_{n} ; \boldsymbol{\mu}_{k}^{(z)}, \boldsymbol{\Sigma}_{k}^{(z)}\big)$ w.r.t. $\boldsymbol{\mu}_{k}^{(z)}$ given by, 
\begin{multline}
\frac{\partial}{\partial \boldsymbol{\mu}_{k}^{(z)}} \mathcal{N}\big(\mathbf{z}_n ;\boldsymbol{\mu}_{k}^{(z)}, \boldsymbol{\Sigma}_{k}^{(z)}\big)
	= -\frac{1}{2}\mathcal{N}\big(\mathbf{z}_n ; \boldsymbol{\mu}_{k}^{(z)}, \boldsymbol{\Sigma}_{k}^{(z)}\big)
	\frac{\partial}{\partial \boldsymbol{\mu}_{k}^{(z)}}\Big\{(\mathbf{z}_n - \boldsymbol{\mu}_{k}^{(z)})^T {\boldsymbol{\Sigma}_{k}^{(z)}}^{-1}(\mathbf{z}_n - \boldsymbol{\mu}_{k}^{(z)})\Big\}
	\label{eq  density derivative w.r.t. mu 2} 
\end{multline}
Moreover, the density derivative  $ $ w.r.t. $ \boldsymbol{\mu}_{k}^{(z)}$is given by 
\begin{equation}
    \frac{\partial}{\partial \boldsymbol{\mu}_{k}^{(z)}}\Big\{(\mathbf{z}_n - \boldsymbol{\mu}_{k}^{(z)})^T {\boldsymbol{\Sigma}_{k}^{(z)}}^{-1}(\mathbf{z}_n - \boldsymbol{\mu}_{k}^{(z)})\Big\}
     =-2(\mathbf{z}_n - \boldsymbol{\mu}_{k}^{(z)})^T{\boldsymbol{\Sigma}_{k}^{(z)}}^{-1}
\end{equation} 


This can simply done by using the matrix rule, $\frac{\partial}{\partial x}\{x^TAx \}=x^TA^T+x^TA $, and setting $ x= \mathbf{z}_n - \boldsymbol{\mu}_{k}^{(z)}, A:= {\mathbf{\Sigma}_k^{(z)}}^{-1}$ and also note $\boldsymbol{\Sigma}_{k}^{(z)}$ is symmetric.

Therefore, the ultimate equation for calculating the expression of $\boldsymbol{\mu}_{k}^{(z)}$ as a function of the responsibility $\gamma$ is derived as follows 
\begin{equation}
\sum_{n=1}^{N}\Bigg[\frac{\frac{e^{-\lambda}\lambda^{k}}{k!}\mathcal{N}\big(\mathbf{z}_{n};\boldsymbol{\mu}_{k}^{(z)},\boldsymbol{\Sigma}_{k}^{(z)}\big)}{\sum_{j=1}^{K}\frac{e^{-\lambda}\lambda^{j}}{j!}\mathcal{N}\big(\mathbf{z}_{n};\mathbf{\mu_{j}},\mathbf{\Sigma_{j}}\big)}
(\mathbf{z}_{n}-\boldsymbol{\mu}_{k}^{(z)})^{T}\{\boldsymbol{\Sigma}_{k}^{(z)}\}^{-1}\Bigg]=0
\end{equation}
since each  $\gamma(t_{n,k})$
being a probability is a scalar, therefore 
$\gamma(t_{n,k})^T=\gamma(t_{n,k})$ gives
\begin{equation}
\boldsymbol{\mu}_{k}^{(z)}=\frac{\sum_{n=1}^{N}\gamma(t_{n,k})\mathbf{z}_n}{\sum_{n=1}^{N}\gamma(t_{n,k})} .
\end{equation}

\subsection{\label{sec:proof:LogLikelihood:MaximizationWRTSigma}Proof of Lemma 2} 

To compute an expression for $\boldsymbol{\Sigma}_{k}^{(z)}$ as a function of responsibility $\gamma$, let us set the partial derivative of log-likelihood w.r.t. $\boldsymbol{\Sigma}_{k}^{(z)}$ by setting zero , given  by
\(\frac{\partial}{\partial \boldsymbol{\Sigma}_{k}^{(z)}} \log f_{\boldsymbol{Z}}(\mathbf{z }; w,\boldsymbol{\mu}, \boldsymbol{\Sigma}) = 0 \). Before compute an expression for $\boldsymbol{\Sigma}_{k}^{(z)}$, it is necessary to simplify the partial derivative  log-likelihood w.r.t. $\boldsymbol{\Sigma}_{k}^{(z)}$  given by
\begin{equation}
\begin{split}
& \frac{\partial}{\partial \boldsymbol{\Sigma}_{k}^{(z)}} \ln f_{\boldsymbol{Z}}(\mathbf{z};w,\boldsymbol{\mu}, \boldsymbol{\Sigma}) \\
& = \frac{\partial}{\partial \boldsymbol{\Sigma}_{k}^{(z)}} \sum_{n=1}^{N}\ln \sum_{k=1}^{K} \frac{e^{-\lambda}\lambda^k}{k!} \mathcal{N}(\mathbf{z}_n;\boldsymbol{\mu}_{k}^{(z)},\boldsymbol{\Sigma}_{k}^{(z)})  = \sum_{n=1}^{N} \frac{\frac{\partial}{\partial \boldsymbol{\Sigma}_{k}^{(z)}} \left( \frac{e^{-\lambda}\lambda^k}{k!} \mathcal{N}(\mathbf{z}_n;\boldsymbol{\mu}_{k}^{(z)},\boldsymbol{\Sigma}_{k}^{(z)}) \right)}{\sum_{k=1}^{K} \frac{e^{-\lambda}\lambda^k}{k!} \mathcal{N}(\mathbf{z}_n;\boldsymbol{\mu}_{k}^{(z)},\boldsymbol{\Sigma}_{k}^{(z)})}, 
\label{eq expression for sigma1}
\end{split}
\end{equation}
as \(\frac{\partial}{\partial \boldsymbol{\Sigma}_{k}^{(z)}}  \frac{e^{-\lambda}\lambda^j}{j!} \mathcal{N}(\mathbf{z}_n;\boldsymbol{\mu}_j,\boldsymbol{\Sigma}_j)=0 \,\forall j\neq k \in {1,\dots,N}.\)

To compute the derivative 
\(\frac{\partial}{\partial \boldsymbol{\Sigma}_{k}^{(z)}}\frac{e^{-\lambda}\lambda^k}{k!} \mathcal{N}(\mathbf{z}_n;\boldsymbol{\mu}_{k}^{(z)},\boldsymbol{\Sigma}_{k}^{(z)})\) in numerator of \eqref{eq expression for sigma1}, we use the logarithmic derivative, expressed as 
\(\frac{\partial}{\partial \boldsymbol{\Sigma}_{k}^{(z)}}\ln \left(\frac{e^{-\lambda}\lambda^k}{k!} \mathcal{N}(\mathbf{z}_n;\boldsymbol{\mu}_{k}^{(z)},\boldsymbol{\Sigma}_{k}^{(z)}) \right)\). This method simplifies the differentiation process by taking advantage of the logarithm's properties to break down the product and exponential components within the function.
Note that 
\begin{multline}
\ensuremath{     \frac{\partial}{\partial \boldsymbol{\Sigma}_{k}^{(z)}}\frac{e^{-\lambda}\lambda^k}{k!} \mathcal{N}\big(\mathbf{z}_n;\boldsymbol{\mu}_{k}^{(z)},\boldsymbol{\Sigma}_{k}^{(z)}\big)} =\frac{e^{-\lambda}\lambda^k}{k!}  \mathcal{N}\big(\mathbf{z}_n;\boldsymbol{\mu}_{k}^{(z)},\boldsymbol{\Sigma}_{k}^{(z)}\big) 
\frac{\partial}{\partial \boldsymbol{\Sigma}_{k}^{(z)}}\ln \Big(\frac{e^{-\lambda}\lambda^k}{k!} \mathcal{N}\big(\mathbf{z}_n;\boldsymbol{\mu}_{k}^{(z)},\boldsymbol{\Sigma}_{k}^{(z)}\big) \Big),
\end{multline} plugging all together gives the partial derivative of log-likelihood w.r.t. $\boldsymbol{\Sigma}_{k}^{(z)}$ as, 

\begin{equation}
\frac{\partial}{\partial\boldsymbol{\Sigma}_{k}^{(z)}}\ln f_{\boldsymbol{Z}}(\mathbf{z};w,\boldsymbol{\mu},\boldsymbol{\Sigma})
=\sum_{n=1}^{N}\gamma(t_{n,k})\frac{\partial}{\partial\boldsymbol{\Sigma}_{k}^{(z)}}\ln\left(\frac{e^{-\lambda}\lambda^{k}}{k!}\mathcal{N}\big(\mathbf{z}_{n};\boldsymbol{\mu}_{k}^{(z)},\boldsymbol{\Sigma}_{k}^{(z)}\big)\right).
\end{equation}

Note here that
\begin{multline}
\ensuremath{\frac{\partial}{\partial\boldsymbol{\Sigma}_{k}^{(z)}}\ln\Big(\frac{e^{-\lambda}\lambda^{k}}{k!}\mathcal{N}\big(\mathbf{z}_{n};\boldsymbol{\mu}_{k}^{(z)},\boldsymbol{\Sigma}_{k}^{(z)}\big)\Big)}
\ensuremath{=\frac{\partial}{\partial\boldsymbol{\Sigma}_{k}^{(z)}}\ln\bigg(\frac{e^{-\lambda}\lambda^{k}}{k!}\bigg)}+\frac{\partial}{\partial\boldsymbol{\Sigma}_{k}^{(z)}}\ln\Big(\frac{1}{(2\pi)^{D/2}|\boldsymbol{\Sigma}_{k}^{(z)}|^{1/2}}\Big)\\
+\frac{\partial}{\partial\boldsymbol{\Sigma}_{k}^{(z)}}\Big\{-\frac{1}{2}(\mathbf{z}_{n}-\boldsymbol{\mu}_{k}^{(z)})^{T}{\boldsymbol{\Sigma}_{k}^{(z)}}^{-1}(\mathbf{z}_{n}-\boldsymbol{\mu}_{k}^{(z)})\Big\}
\end{multline}
For the first part, $\ln \big (\frac{e^{-\lambda}\lambda^k}{k!}\big)$ is constant w.r.t. $\boldsymbol{\Sigma}_{k}^{(z)}$, hence $\frac{\partial}{\partial \boldsymbol{\Sigma}_{k}^{(z)}} \Big\{ \ln \frac{e^{-\lambda}\lambda^k}{k!} \Big\}=0$. For the second part, 
\begin{equation}\frac{\partial}{\partial \boldsymbol{\Sigma}_{k}^{(z)}} \Big\{ \ln \Big(\frac{1}{(2\pi)^{D/2}|\boldsymbol{\Sigma}_{k}^{(z)}|^{1/2}}\Big)\Big\}
= -\frac{1}{2}{\boldsymbol{\Sigma}_{k}^{(z)}}^{-1} \end{equation}
This can be computed using the matrix rule  
$\frac{\partial}{\partial  \boldsymbol{A}} |\boldsymbol{A}|=| \boldsymbol{A}|( \boldsymbol{A}^{-1})^T;$
and note $\boldsymbol{\Sigma}_{k}^{(z)}$ is symmetric. For the third part, the partial derivative of $\frac{1}{2}(\mathbf{z}_n -\boldsymbol{\mu}_{k}^{(z)})^T{\boldsymbol{\Sigma}_{k}^{(z)}}^{-1}(\mathbf{z}_n - \boldsymbol{\mu}_{k}^{(z)})$ w.r.t. $\boldsymbol{\Sigma}_{k}^{(z)}$
can be computed using the matrix rule  \(\frac{\partial}{\partial \mathbf{A}} (\mathbf{x^TA^{-1}x})=\mathbf{-A^{-1}xx^TA^{-1}},\)
considering, \(\xspace \mathbf{x}=\mathbf{z}_n -\boldsymbol{\mu}_{k}^{(z)} ; \mathbf{A}= \boldsymbol{\Sigma}_{k}^{(z)}\), which is given by
\begin{multline}
    \frac{\partial}{\partial \boldsymbol{\Sigma}_{k}^{(z)}} \Big\{-\frac{1}{2}(\mathbf{z}_n -\boldsymbol{\mu}_{k}^{(z)})^T{\boldsymbol{\Sigma}_{k}^{(z)}}^{-1}(\mathbf{z}_n - \boldsymbol{\mu}_{k}^{(z)})\Big\}
 = -\frac{1}{2}\frac{\partial}{\partial \boldsymbol{\Sigma}_{k}^{(z)}} \big\{(\mathbf{z}_n -\boldsymbol{\mu}_{k}^{(z)})^T{\boldsymbol{\Sigma}_{k}^{(z)}}^{-1}(\mathbf{z}_n - \boldsymbol{\mu}_{k}^{(z)})\big\}\\
=\frac{1}{2}{\boldsymbol{\Sigma}_{k}^{(z)}}^{-1}(\mathbf{z}_n - \boldsymbol{\mu}_{k}^{(z)})(\mathbf{z}_n -\boldsymbol{\mu}_{k}^{(z)})^T{\boldsymbol{\Sigma}_{k}^{(z)}}^{-1}.
\end{multline}
Therefore, we have
\begin{multline}
	\frac{\partial}{\partial \boldsymbol{\Sigma}_{k}^{(z)}} \ln\Big( \frac{e^{-\lambda}\lambda^k}{k!} \mathcal{N}(\mathbf{z}_n;\boldsymbol{\mu}_{k}^{(z)},\mathbf{\boldsymbol{\Sigma}_{k}^{(z)}}) \Big)
	=-\frac{1}{2}{\boldsymbol{\Sigma}_{k}^{(z)}}^{-1} + \frac{1}{2}{\boldsymbol{\Sigma}_{k}^{(z)}}^{-1}(\mathbf{z}_n - \boldsymbol{\mu}_{k}^{(z)})(\mathbf{z}_n -\boldsymbol{\mu}_{k}^{(z)})^T{\boldsymbol{\Sigma}_{k}^{(z)}}^{-1}.
\end{multline}
Hence, the final expression of the partial derivative of log-likelihood w.r.t. $\boldsymbol{\Sigma}_{k}^{(z)}$ by
\begin{multline}
\frac{\partial}{\partial\boldsymbol{\Sigma}_{k}^{(z)}}\ln f_{\boldsymbol{Z}}(\mathbf{z};w,\boldsymbol{\mu},\boldsymbol{\Sigma})
=\sum_{n=1}^{N}\gamma(t_{n,k}) \Bigg\{ -\frac{1}{2}{\boldsymbol{\Sigma}_{k}^{(z)}}^{-1}+ \frac{1}{2}{\boldsymbol{\Sigma}_{k}^{(z)}}^{-1}(\mathbf{z}_n - \boldsymbol{\mu}_{k}^{(z)})(\mathbf{z}_n -\boldsymbol{\mu}_{k}^{(z)})^T{\boldsymbol{\Sigma}_{k}^{(z)}}^{-1}\Bigg\}.
\end{multline}
Finally, maximizing log-likelihood w.r.t. $\boldsymbol{\Sigma}_{k}^{(z)}$ by setting the derivative zero gives 
\begin{equation}
\sum_{n=1}^{N}\gamma(t_{n,k})\{\boldsymbol{\Sigma}_{k}^{(z)}\}^{-1}
\left[(\mathbf{z}_{n}-\mu_{k}^{(z)}) (\mathbf{z}_{n}-\mu_{k}^{(z)})^{T}\{\boldsymbol{\Sigma}_{k}^{(z)}\}^{-1}-1 \right]=0.
\end{equation}
Therefore we have the expression for $\boldsymbol{\Sigma}_{k}^{(z)}$ depending on the responsibility $\gamma$ as,
\begin{equation}
    \boldsymbol{\Sigma}_{k}^{(z)} = \frac{\sum_{n=1}^{N} \gamma(t_{n,k}) (\mathbf{z}_n - \mu_k^{(z)}) (\mathbf{z}_n - \mu_k^{(z)})^T}{\sum_{i=1}^{N} \gamma(t_{n,k})}. 
\end{equation}

{\subsection{\label{sec:proof:} The conditional density \(f_{\boldsymbol{Z}}(\boldsymbol{z} | t_k = 0)\) given that \(\boldsymbol{z}\) does not belong to the \(k\)-th Gaussian component}

To simplify the calculation for \( f_{\boldsymbol{Z}}(\boldsymbol{z} | t_k = 0) \), let's break down the steps in a more straightforward way. The \ac{GMM} for the noise \( \boldsymbol{Z} \) is given by \eqref{eq approx pdf of joint quantum noise in Gaussian form for vector}
   The conditional density when a data point belongs to the \( k \)-th Gaussian component is given in Section III(A).
   The joint probability is computed in \eqref{eq The joint probability}.
   The conditional density the density of \( \boldsymbol{z} \) given that it does not belong to the \( k \)-th Gaussian component is given by \( f_{\boldsymbol{Z}}(\boldsymbol{z} | t_k = 0) \). It can be calculated by considering all components except the \( k \)-th
   \begin{equation}
        f_{\boldsymbol{Z}}(\boldsymbol{z} | t_k = 0) = \frac{f_{\boldsymbol{Z}}(\boldsymbol{z}) - f_{\boldsymbol{Z}}(\boldsymbol{z}, t_k = 1)}{1 - f_{\boldsymbol{Z}}(t_k = 1)}.
   \label{}
   \end{equation}
   Using the expressions for \( f_{\boldsymbol{Z}}(\boldsymbol{z}) \) and \( f_{\boldsymbol{Z}}(\boldsymbol{z}, t_k = 1) \),
   \begin{equation}
       f_{\boldsymbol{Z}}(\boldsymbol{z} | t_k = 0) = \frac{\sum_{i=0}^{K} \frac{e^{-\lambda} \lambda^i}{i!} N(\boldsymbol{z}; \mu_i^{(\boldsymbol{z})}, \Sigma_i^{(\boldsymbol{z})}) - \frac{e^{-\lambda} \lambda^k}{k!} N(\boldsymbol{z}; \mu_k^{(\boldsymbol{z})}, \Sigma_k^{(\boldsymbol{z})})}{1 - \frac{e^{-\lambda} \lambda^k}{k!}}.
   \end{equation}
   The numerator simplifies to
   \(
   \sum_{\substack{i=0 \\ i \neq k}}^{K} \frac{e^{-\lambda} \lambda^i}{i!} N(\boldsymbol{z}; \mu_i^{(\boldsymbol{z})}, \Sigma_i^{(\boldsymbol{z})}), 
   \)
   and the denominator becomes
   \(
   1 - \frac{e^{-\lambda} \lambda^k}{k!}\), that is \( \sum_{\substack{i=0 \\ i \neq k}}^{K} \frac{e^{-\lambda} \lambda^i}{i!}.
   \)
   Therefore, the simplified expression for \( f_{\boldsymbol{Z}}(\boldsymbol{z} | t_k = 0) \) is
   \begin{equation}
       f_{\boldsymbol{Z}}(\boldsymbol{z} | t_k = 0) = \biggl[\sum\limits_{\substack{i=0 \\ i \neq k}}^{K} \frac{e^{-\lambda} \lambda^i}{i!} N(\boldsymbol{z}; \mu_i^{(\boldsymbol{z})}, \Sigma_i^{(\boldsymbol{z})})\biggl]/\biggl[\sum\limits_{\substack{i=0 \\ i \neq k}}^{K} \frac{e^{-\lambda} \lambda^i}{i!}\biggl].
   \label{}
   \end{equation}
This formula provides the \ac{p.d.f.} for \( \boldsymbol{z} \) given that it does not belong to the \( k \)-th component, expressed as a weighted sum of all other Gaussian components.}


\begin{acronym}
     \acro{LOQC}{linear optical quantum computing}
    \acro{ML}{Machine Learning}
    \acro{MAP}{Maximum a-Posteriori}
    \acro{DL}{Deep Learning}
    \acro{RL}{Reinforcement Learning}
    \acro{SVM}{Support Vector Machines}
    \acro{PCA}{Principal Component Analysis}
    \acro{KF}{Kalman Filter}
    \acro{QML}{Quantum Machine Learning}
    \acro{NN}{Neural Network}
    \acro{EM}{Expectation-Maximization}
    \acro{GM}{Gaussian Mixture}

\acro{WCSS}{within-cluster sum of squares }
\acro{BIC}{Bayesian Information Criterion }
 \acro{AIC}{Akaike Information Criterion}   
    \acro{AWGN}{Additive-White-Gaussian Noise}
    \acro{SKR}{Secret Key Rate}
    \acro{QKD}{Quantum Key Distribution}
    \acro{PNS}{Photon Number Splitting}
    \acro{CV-QKD}{Continuous-Variable QKD}
    \acro{FSO}{Free-Space Optics}
    \acro{MDI}{Measure-device-independent}
    \acro{DV-QKD}{Discrete-Variable QKD}
    \acro{CV}{continuous variables}
    \acro{DV}{discrete variables}
    \acro{CPTP}{completely positive, trace preserving}
    \acro{PM}{Prepare-and-Measure}
    \acro{PDTC}{Probability Distribution of Transmission Coefficient}
    \acro{S/C}{spacecraft}
    \acro{p.d.f.}{probability density function}
    \acro{p.m.f.}{probability mass function}
    \acro{SNR}{Signal-to-Noise Ratio}
    \acro{SPS}{Single-Source Photon}
    \acro{GMM}{Gaussian Mixture Model}
    \acro{GMs}{Gaussian Mixtures}
    \acro{GS}{Ground Station}
\end{acronym}
\bibliographystyle{IEEEtran}

\bibliography{IEEEabrv,paper}

\end{document}